\documentclass[pra,twocolumn,amsmath,amssymb,superscriptaddress]{revtex4-2}

\usepackage{epsfig,amsmath}
\usepackage{subfigure}
\usepackage{graphicx}% Include figure files
\usepackage{dcolumn}% Align table columns on decimal point
\usepackage{stmaryrd}
\usepackage{mathrsfs}
\usepackage{pifont}
\usepackage{amsmath}
\usepackage{amssymb}
\usepackage{bm}
\usepackage{latexsym}
\usepackage{color}
\usepackage{epstopdf}
\usepackage{upgreek}
\usepackage{times}
\usepackage[colorlinks=true,linkcolor=blue,citecolor=blue,urlcolor=blue]{hyperref}
\begin{document}
\title{Remote magnon-phonon entanglement in waveguide magnomechanics}
\author{Shi-fan Qi}
\email{qishifan@hebtu.edu.cn}
\affiliation{College of Physics and Hebei Key Laboratory of Photophysics Research and Application, Hebei Normal University, Shijiazhuang 050024, China}

\author{Fan Li}
\affiliation{College of Physics and Hebei Key Laboratory of Photophysics Research and Application, Hebei Normal University, Shijiazhuang 050024, China}

\begin{abstract}
Generating long-distance quantum entanglement is crucial for advancing quantum information processing. In this work, we propose a protocol for generating remote magnon-phonon entanglement in a hybrid waveguide magnomechanical system, where multiple spatially separated magnon modes couple to a common waveguide while interacting with their respective phonon modes. By applying tailored pulsed drives and engineering the magnomechanical interactions, our scheme enables the creation of diverse long-distance and dynamically stable entanglement. Beyond basic magnon-phonon two-mode entanglement, it supports genuine multimode entanglement between a single phonon and multiple magnons, bipartite entanglement between a single magnon and multiple phonons, as well as genuine four-mode entanglement involving two magnons and two phonons. Moreover, we show that dissipative magnon-magnon interactions mediated by traveling photons can generate substantially stronger remote entanglement than coherent couplings. Our work provides an experimentally feasible scheme for the remote generation of magnon-phonon entanglement.
\end{abstract}

\maketitle
\section{Introduction}
Quantum entanglement between spatially separated systems is a vital resource for diverse quantum technologies~\cite{Remoteent}, underpinning quantum networks~\cite{quantint,quantint2}, quantum computing~\cite{quantumcomputing}, and quantum communication~\cite{quantumcommunication}. Significant progress has been made in generating and distributing remote entanglement across various quantum platforms~\cite{stablerement,teleportmicro,remotegiantatom,remotespin,remotefiber,distantgate}, yet its efficient realization remains a central and active topic of current research~\cite{remotenanocavity,remotequbit,remotemax,remotesquee,remoteQED,remoteoptomech}. Extending beyond bipartite correlations, multipartite entanglement~\cite{decohenmulent,multientnet,multioptimiza} serves as a key enabling ingredient for diverse quantum-networking protocols, including quantum state sharing~\cite{stateshar}, quantum teleportation~\cite{multitelepor,teleportation}, and quantum secret sharing~\cite{secret}. Accordingly, it is of considerable importance to develop experimentally feasible schemes for preparing remote multipartite entanglement within specific physical platforms~\cite{multimicrowave,optcavity,circuit,squemulent,ionmulent}.

Hybrid quantum systems based on magnons in ferromagnetic crystals have attracted considerable interest owing to the long coherence time of spin ensembles~\cite{cavitymagnonics,quantummagnon}, thereby providing a promising platform for entanglement generation and applications~\cite{remoteschro,magquantifi,tunable,nonreciproopto,Nonrephophon,tricoupling,tailoring,Barnett}. In particular, cavity magnomechanical systems~\cite{magnoncavity,kerrcavitymagnon,cavitymagnomechan}, in which a yttrium iron garnet (YIG) sphere is placed inside a microwave cavity, enable the magnon mode to couple to photons via magnetic dipole interactions and to phonons via magnetostrictive forces. Building on this platform, various schemes have been proposed for generating stable continuous-variable entanglement~\cite{nonreciprocalent,entopa,kerrmagnontmss,macroent}, including genuine tripartite photon-magnon-phonon entanglement~\cite{mppentang} and optical-magnon-microwave-phonon entanglement~\cite{COMMLJ}. Furthermore, the hybrid cavity-magnon-qubit system~\cite{magnonqubit,qubitmagnon,qubitmagnonsquee} provides an alternative route for realizing discrete-variable entanglement, such as photon-magnon-qubit GHZ states~\cite{GHZ} and magnonic Bell~\cite{Bell} and NOON states~\cite{NOON}. However, these schemes are limited to local configurations and cannot generate remote entanglement. Recently, it has been shown that propagating microwave photons can mediate remote magnon-magnon coupling~\cite{coherremag}, which can be exploited to generate remote bipartite magnon-magnon entanglement~\cite{Remagnonent,remotemagnon2}.

The waveguide magnon system~\cite{Anomagnon,giantspin,SingleMode,Unidirectional} can induce long-distance magnon-magnon coupling, but it is generally ineffective for generating magnonic entanglement without external squeezing fields. To address this limitation and enable entanglement across different frequency modes, we combine the waveguide magnon systems~\cite{Unidirectional} with magnomechanical systems~\cite{magnoncavity}, forming a new hybrid waveguide
magnomechanical architecture. In this work, we analytically and numerically investigate the generation of entanglement between distant subsystems. By applying tailored driving fields and tuning the magnomechanical interactions, we realize several types of remote magnon-phonon entanglement, including two-mode magnon-phonon entanglement, genuine multimode entanglement between a single phonon and multiple magnon modes, multimode entanglement between a single magnon and multiple phonon modes, as well as genuine four-mode entanglement in a two-magnon-two-phonon system. Our results reveal that waveguide-mediated magnon-magnon interactions play a crucial role in generating the entanglement, with dissipative coupling outperforming coherent coupling for a given interaction strength.

The paper is organized as follows. In Sec.~\ref{Secmodel}, we introduce thewaveguide magnomechanical system and present the corresponding Hamiltonian. The system dynamics and the criteria to quantify entanglement are provided in Sec.~\ref{secentcriterion}. Section~\ref{secresult} presents the results on the generation of remote two-mode magnon-phonon entanglement, multimode entanglement involving a single magnon and multiple phonons or a single phonon and multiple magnons, as well as genuine four-mode entanglement in dual magnon-phonon systems. Finally, discussions and conclusions are provided in Sec.~\ref{Secconclu}. 

\section{Thewaveguide magnomechanical system}\label{Secmodel}
\begin{figure}[t]
	\centering
	\includegraphics[width=0.48\textwidth]{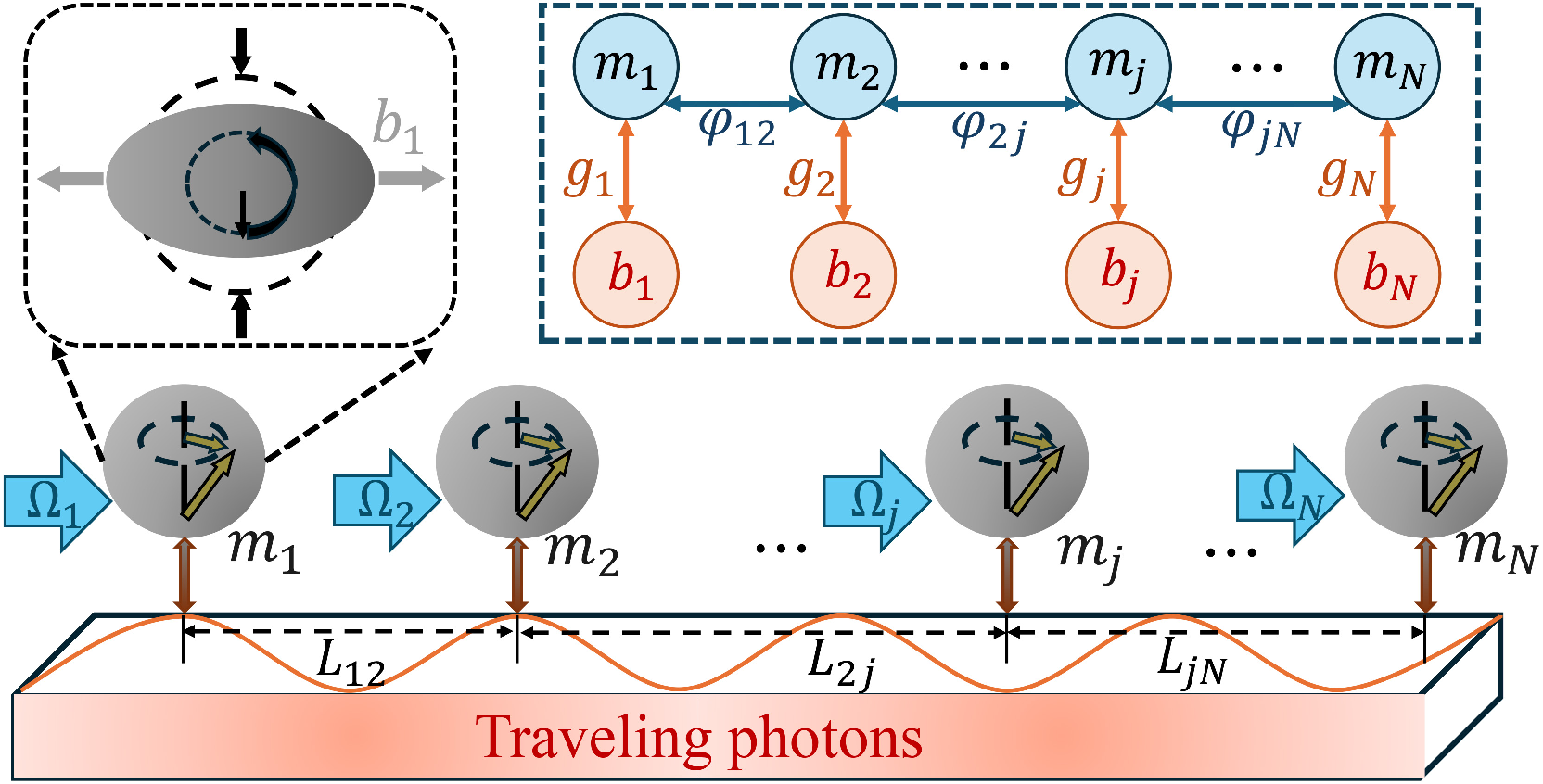}
	\caption{Schematic diagram: $N$ small YIG spheres (magnon modes) are coupled to a common waveguide, and the effective interactions between magnon modes $m_j$ and $m_l$ are mediated by traveling photons propagating in the waveguide. Moreover, the magnon mode $m_j$ in each YIG sphere is driven by a strong control field with Rabi frequency $\Omega_j$. The upper left inset illustrates the magnetostrictive mechanism, where dynamic magnon magnetization (represented by vertical black arrows) induces mechanical deformation in the YIG sphere, thereby coupling to its phonon mode $b_j$. The right panel schematically depicts the mode couplings, where the magnon mode $m_j$ couples to its phonon mode with strength $g_j$, while $\varphi_{jl}$ denotes the quantum traveling phase between magnon modes $m_j$ and $m_l$, which determines the form of the corresponding interaction.}\label{diagram}
\end{figure}
As depicted in Fig.~\ref{diagram}, the hybrid quantum system under consideration consists of $N$ YIG spheres (which provide magnon modes) coupled to a microstrip waveguide at distinct locations~\cite{giantspin}. The magnon modes in different spheres interact with waveguide photon modes via the Zeeman interaction and with their respective phonon modes via the magnetostrictive interaction, where each phonon mode corresponds to the mechanical deformation of the YIG sphere~\cite{magnoncavity,kerrcavitymagnon,cavitymagnomechan}. The system Hamiltonian can be written as
\begin{equation}\label{Htotorigin}
\begin{aligned}
	H_{\rm tot}&=H_s+H_I+H_d,\\
    H_s&=\sum^N_{j=1}\omega_jm^\dag_jm_j+\omega_bb^\dag_jb_j+\tilde{g}_jm^\dag_jm_j(b_j+b^\dag_j),\\
	H_I&=-i\sum^{N-1}_{j=1}\sum^N_{l=j+1}\sqrt{\kappa_j\kappa_l}e^{i\varphi_{jl}}(m^\dag_jm_l+m_jm^\dag_l),\\
	H_d&=\sum^N_{j=1}\Omega_j(e^{-i\epsilon t}m^\dag_j+e^{i\epsilon t}m_j),
\end{aligned}
\end{equation}
where $H_s$ represents the Hamiltonian of $N$ magnomechanical systems, $H_I$ denotes the interaction between magnon modes, and $H_d$ describes the driving Hamiltonian. $\omega_j$ is the transition frequency of magnon mode $m_j$, which is given by $\omega_j=\tilde{\gamma}h_j$, where $\tilde{\gamma}$ is the gyromagnetic ratio and $h_j$ is the external bias magnetic field, allowing for tuning via the external magnetic field. $\omega_b$ denotes the transition frequency of the phonon modes, where the frequencies of phonon modes in different YIG spheres are assumed to be identical. $\tilde{g}_j$ represents the single-excitation magnomechanical coupling strength, which can be tuned from $0$ to $60$ mHz by adjusting the direction of the bias field in recent experiments~\cite{magnoncavity}. The coupling between magnon modes is mediated by traveling photons in the waveguide and consists of coherent and dissipative components, given by $\sqrt{\kappa_j\kappa_l}\sin{\varphi_{jl}}$ and $\sqrt{\kappa_j\kappa_l}\cos{\varphi_{jl}}$, respectively. Here, $\varphi_{jl}$ denotes the propagation phase accumulated by traveling photons over a distance $L_{jl}$, while $\kappa_j$ and $\kappa_l$ are the radiative decay rates of magnons $m_j$ and $m_l$ induced by the waveguide photons (see Appendix~\ref{appa} for details). Owing to the periodicity $e^{i(2k\pi+\varphi_{jl})}=e^{i\varphi_{jl}}$ for any integer $k$, the phase is defined modulo $2\pi$, and the results remain invariant under shifts by integer multiples of $2\pi$. The Rabi frequency, $\Omega_j=\tilde{\gamma}\sqrt{5\bar{N}}B_j/4$, characterizes the amplitude of the external driving field with frequency $\epsilon$, where $B_j$ is the magnetic field strength of the driving field and $\bar{N}$ represents the total number of spins in the YIG sphere~\cite{cavitymagnonics}.

Under strong-driving fields, the system's dynamics can be linearized by writing the operators $o_j=\langle o_j\rangle +\delta o_j, o=m,b$, where $\langle o_j\rangle$ is the steady amplitude of mode $o_j$ and $\delta o_j$ is the corresponding fluctuation operator. After the standard linearization approximation~\cite{optcavity}, the system Hamiltonian in Eq.~\eqref{Htotorigin} turns into
\begin{equation}\label{Hamsystem}
\begin{aligned}
	H&=\sum_{j=1}^N\Delta_jm^\dag_jm_j+\omega_bb^\dag_jb_j+g_j(m_j+m^\dag_j)(b_j+b^\dag_j)\\
	&\quad -i\sum^{N-1}_{j=1}\sum^N_{l=j+1}\sqrt{\kappa_j\kappa_l}e^{i\varphi_{jl}}(m^\dag_jm_l+m_jm^\dag_l),\\
\end{aligned}
\end{equation}
where $\Delta_j$ is the effective detuning of magnon mode $m_j$ and $g_j=\tilde{g}_j\langle m_j\rangle$ is the driving-enhanced magnomechanical coupling strength. Here, for simplicity, we apply the substitution $\delta o_j \to o_j$. Further details are provided in Appendix~\ref{appb}.

\section{System dynamics and Entanglement criterion}\label{secentcriterion}
In the framework of open quantum systems, under standard Markovian approximations and employing the quantum Langevin equation (QLE), the system dynamics governed by the Hamiltonian in Eq.~\eqref{Hamsystem} can be expressed in matrix form
\begin{equation}\label{dotuwhole}
	\dot{u}(t)=Au(t)+\epsilon^{\rm in}(t).
\end{equation}
$u(t)$ is an operator vector, with components are given by $u_{4j-3}=X_{m_j}$, $u_{4j-2}=Y_{m_j}$, $u_{4j-1}=X_{b_j}$, and $u_{4j}=Y_{b_j}$ for $j=1,2,\cdots N$, where $u_j$ denotes the $j$-th component of $u(t)$. $X_{o_j}=(o_j+o^\dag_j)/\sqrt{2}$ and $Y_{o_j}=(o_j-o^\dag_j)/i\sqrt{2}$ are the quadrature operators of mode $o_j$ with $o=m,b$. The drift matrix $A$ is a square matrix of size $4N\times 4N$, whose elements can be expressed as
\begin{equation}\label{driftANmag}
\begin{aligned}
	A_{4j-3,4j-2}&=-A_{4j-2,4j-3}=\Delta_j,\\
	A_{4j-1,4j}&=-A_{4j,4j-1}=\omega_b,\\
	A_{4j-3,4j-3}&=A_{4j-2,4j-2}=-\tilde{\kappa}_j,\\
	A_{4j-1,4j-1}&=A_{4j,4j}=-\kappa_b,\\
	A_{4j-2,4j-1}&=A_{4j,4j-3}=-2g_j,\\
	A_{4j-3,4l-3}&=A_{4j-2,4l-2}=-\sqrt{\kappa_j\kappa_l}\cos\varphi_{jl}, \quad j<l,\\
	A_{4j-3,4l-2}&=-A_{4j-2,4l-3}=\sqrt{\kappa_j\kappa_l}\sin\varphi_{jl},\quad j<l,
\end{aligned}
\end{equation}
where $A_{jl}$ denotes the matrix elements in the $j$-th row and $l$-th column. $\tilde{\kappa}_j\equiv\kappa_j+\gamma$, with $\gamma$ being the intrinsic decay rate of magnon mode. $\epsilon^{\rm in}(t)$ is the vector of Gaussian noise operators, its elements are defined as
$\epsilon_{4j-3}^{\rm in}=\tilde{X}_j^{\rm in}$, $\epsilon_{4j-2}^{\rm in}=\tilde{Y}_j^{\rm in}$,
$\epsilon_{4j-1}^{\rm in}=X_{b_j}^{\rm in}$, $\epsilon_{4j}^{\rm in}=Y_{b_j}^{\rm in}$, with
$\tilde{X}_j^{\rm in}\equiv\sqrt{2\tilde{\kappa}_j}X^{\rm in}_{m_j}+\sum^N_{l\neq j}\sqrt{2\Gamma_{jl}}X^{\rm in}_{m_l}$
and $\tilde{Y}_j^{\rm in}\equiv\sqrt{2\tilde{\kappa}_j}Y^{\rm in}_{m_j}+\sum^N_{l\neq j}\sqrt{2\Gamma_{jl}}Y^{\rm in}_{m_l}$, where $\Gamma_{jl}\equiv\sqrt{\kappa_j\kappa_l}\cos\varphi_{jl}$. The quadrature noise components are defined as $X_{o_j}^{\rm in}=({o_j}_{\rm in}+o^\dag_{j_{\rm in}})/\sqrt{2}$ and $Y_{o_j}^{\rm in}=({o_j}_{\rm in}-o^\dag_{j_{\rm in}})/i\sqrt{2}$. $m_{j_{\rm in}}$ and $b_{j_{\rm in}}$ are the input noise operators for $m_j$ and $b_j$, respectively, characterized by their corresponding covariance functions: $\langle m^\dag_{j_{\rm in}}(t)m_{j_{\rm in}}(t')\rangle=\bar{n}\delta(t-t')$ and $\langle b^\dag_{j_{\rm in}}(t)b_{j_{\rm in}}(t')\rangle=\bar{n}_{b}\delta(t-t')$. $\bar{n}$ and $\bar{n}_b$ represent the mean populations at thermal equilibrium for the magnon and phonon modes, respectively. The derivations are provided in Appendix~\ref{appb} and Eq.~\eqref{dotuwhole} is the matrix form of Eq.~\eqref{deltaoper}.

Given the condition that all modes are initially occupied at their respective thermal states, the dynamics of the hybrid system can be fully described by the covariance matrix (CM) $V(t)$ with dimension $4N\times4N$. Via the QLE shown in Eq.~\eqref{dotuwhole}, the CM $V(t)$ satisfies
\begin{equation}\label{cmvt}
	\dot{V}(t)=AV(t)+V(t)A^T+D,
\end{equation}
where the elements of $V$ are defined as $V_{jl}=\langle u_j(t)u_l(t)+u_l(t)u_j(t)\rangle/2$. The diffusion matrix $D$ can be derived as
\begin{equation}\label{driftDNmag}
\begin{aligned}
	D_{4j-3,4j-3}&=D_{4j-2,4j-2}=\tilde{\kappa}_j(2\bar{n}+1),\\
	D_{4j-1,4j-1}&=D_{4j,4j}=\kappa_b(2\bar{n}_b+1),\\
	D_{4j-3,4l-3}&=D_{4j-2,4l-2}=\Gamma_{jl}(2\bar{n}+1),~j<l,	 
\end{aligned}
\end{equation}	
via its definitions $D_{jl}=\langle\epsilon^{\rm in}_j(t)\epsilon_l^{\rm in}(t)+\epsilon^{\rm in}_l(t)\epsilon_j^{\rm in}(t)\rangle/2$. With the initial condition, $V_{4j-3,4j-3}(0)=V_{4j-2,4j-2}(0)=\bar{n}+1/2$ and $V_{4j-1,4j-1}(0)=V_{4j,4j}(0)=\bar{n}_b+1/2$, the CM $V(t)$ can be obtained numerically by Eq.~\eqref{cmvt}.

The logarithmic negativity (LN)~\cite{computable,quantscaling,LNmonotone,logarithmic} is adopted to quantify the bipartite quantum entanglement, with larger LN values indicating stronger entanglement. For Gaussian states of modes $o$ and $p$ in the continuous-variable system, the LN can be defined in terms of the CM, which simplifies the treatment of the infinite-dimensional density matrix and facilitates practical calculations, as~\cite{quantscaling}
\begin{equation}\label{lognegayivity}
	E_{o|p}={\rm max}[0,-{\rm ln}(2\tilde{e}_{o|p})],
\end{equation}
where $\tilde{e}_{o|p}={\rm min}\{{\rm eig}|i\sigma_2\tilde{V}_4|\}$ (with the symplectic matrix $\sigma_2=\oplus^2_{j=1}i\sigma_y$ and $\sigma_y$ is the $y$-Pauli matrix) is the minimum symplectic eigenvalue of the CM $\tilde{V}_4=P_{o|p}V_4P_{o|p}$. Here, $\tilde{V}_4$ is the $4\times4$ CM of two-mode subsystems, obtained by removing the rows and columns corresponding to the uninteresting modes of $V$, and $P_{o|p}=Diag[1,-1,1,1]$ is the matrix that realizes partial transposition at $V_4$. To calculate the one-mode-versus-multi-mode LN, $E_{o|p_1p_2\cdots p_n}$, one simply follows the definition in Eq.~\eqref{lognegayivity}, replacing $\sigma_2=\bigoplus_{j=1}^{2} i\sigma_y$ with $\sigma_n=\bigoplus_{j=1}^{n}i\sigma_y$, and replacing $\tilde{V}_4 = P_{o|p}V_4P_{o|p}$ with $\tilde{V} = P_{o|p_1p_2\cdots p_n} V P_{o|p_1p_2\cdots p_n}$, where $P_{o|p_1p_2\cdots p_n}=Diag[1,-1,1,1\cdots 1,1]$ is the partial transposition matrix with dimension $2n+2$. Similarly, for the two-mode-vs-two-mode LN $E_{o_1o_2|p_1p_2}$, the partial transposition matrix becomes $P_{o_1o_2|p_1p_2}=Diag[1,-1,1,-1,1,1,1,1]$ and the LN is given by $E_{o_1o_2|p_1p_2}=\sum_j{\rm max}[0,-{\rm ln}(2\tilde{e}_j)]$, where $\tilde{e}_j={\rm eig}|i\sigma_4\tilde{V}_8|$ is the symplectic eigenvalue of the CM $\tilde{V}_8=P_{o_1o_2|p_1p_2}V_8P_{o_1o_2|p_1p_2}$. Consequently, one can obtain the time-dependent LNs for all the above bipartitions once the system CM is numerically calculated using Eq.~\eqref{cmvt}. 

\section{Numerical results about remote magnon-phonon entanglement}\label{secresult}
The preceding analysis provides a theoretical framework for investigating entanglement generation in this hybrid waveguide magnomechanical system. We now apply it to concrete models to realize remote magnon-phonon two-mode entanglement, multimode entanglement, and genuine four-mode entanglement.

\subsection{Two-mode entanglement}
\begin{figure}[htbp]
	\centering
	\includegraphics[width=0.48\textwidth]{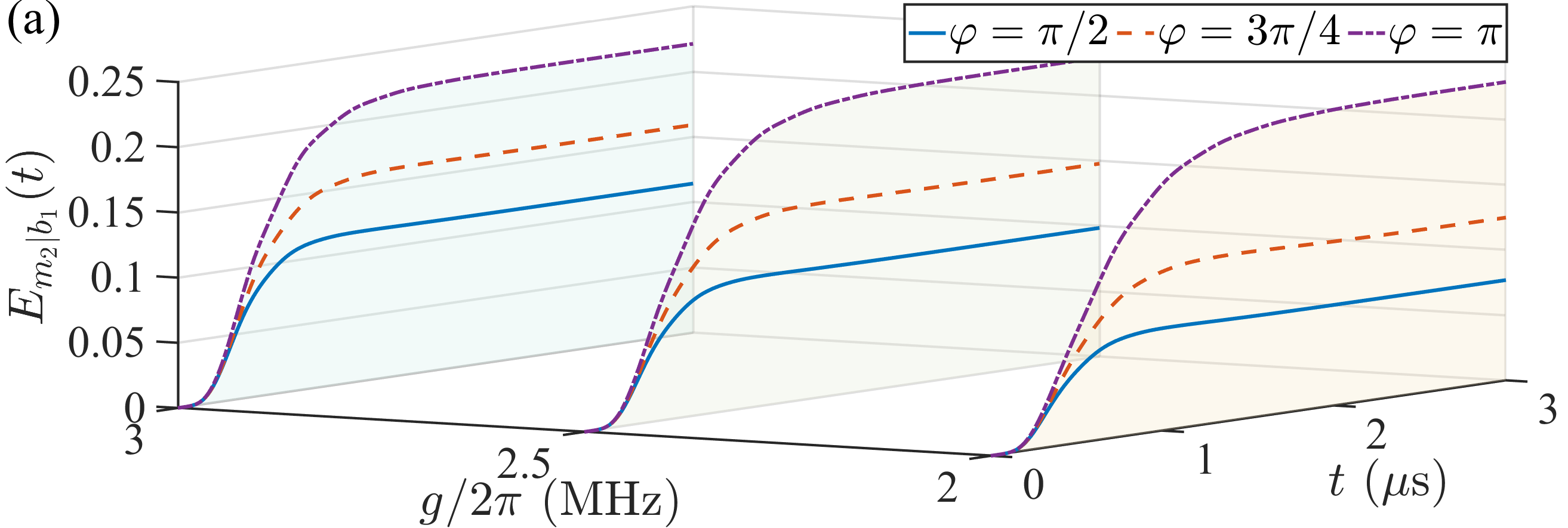}
	\includegraphics[width=0.235\textwidth]{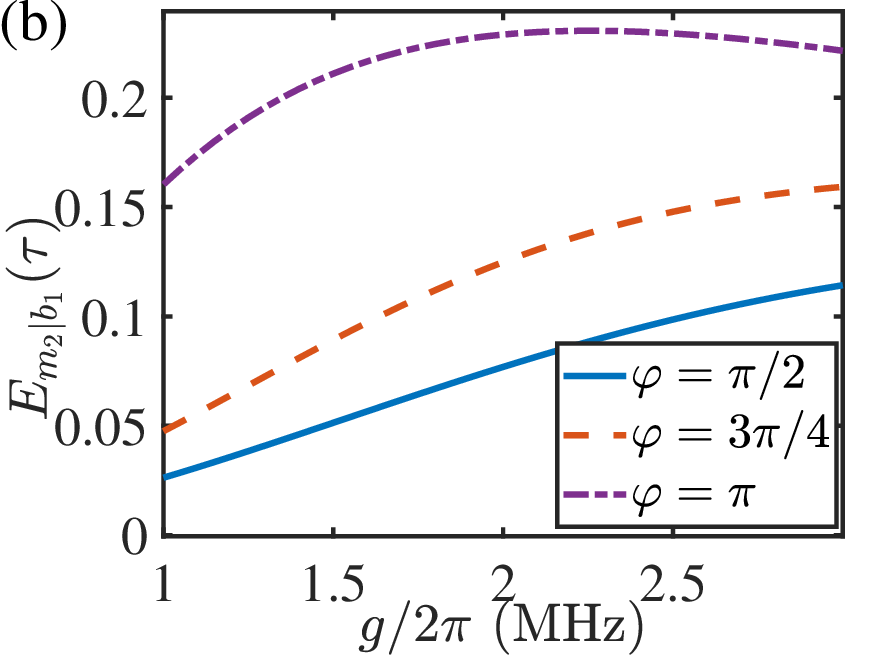}
	\includegraphics[width=0.235\textwidth]{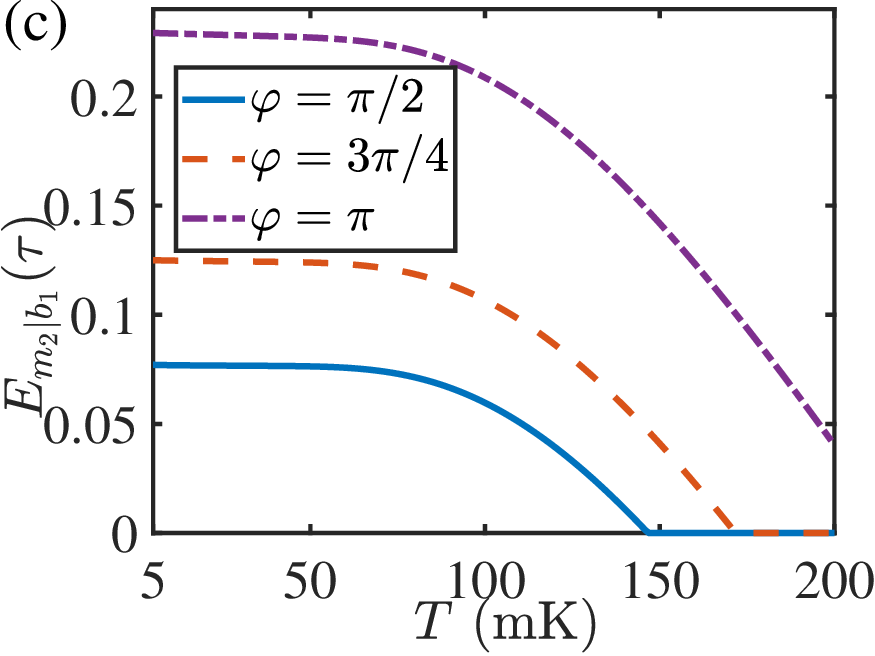}
	\includegraphics[width=0.235\textwidth]{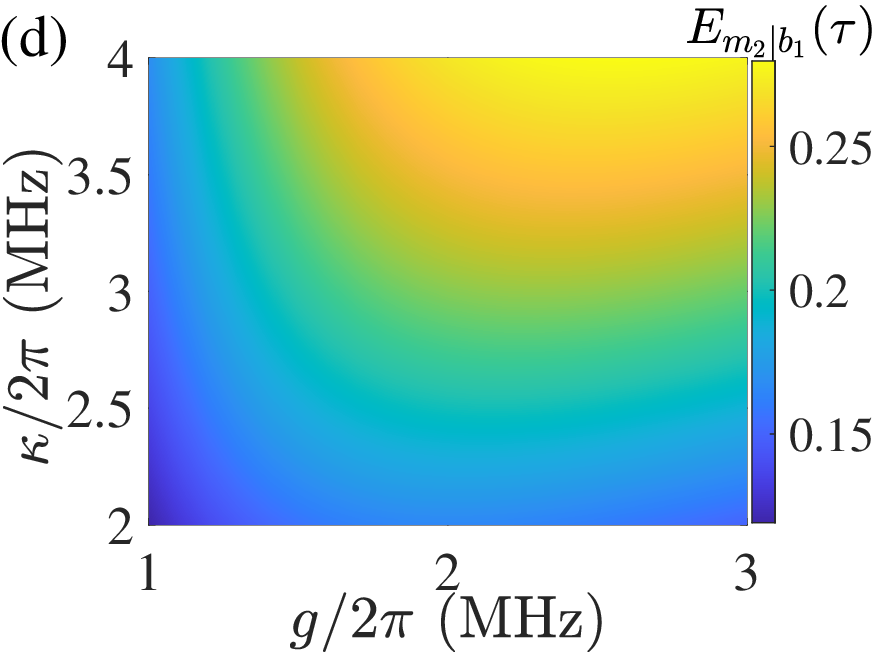}
	\includegraphics[width=0.235\textwidth]{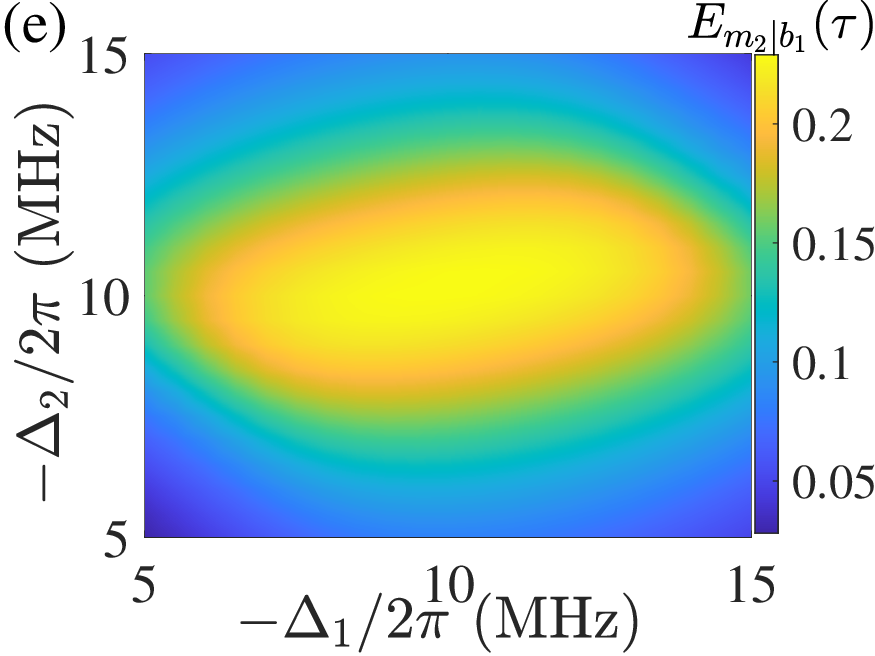}
	\caption{(a) Time evolution of the LN quantifying bipartite entanglement between modes $m_2$ and $b_1$, denoted as $E_{m_2|b_1}(t)$, under varying driving-enhanced magnomechanical coupling strength $g$ and quantum phase $\varphi$. (b) LN $E_{m_2|b_1}(\tau)$ at a fixed time $\tau$ as a function of $g$ under different values of $\varphi$. (c) $E_{m_2|b_1}(\tau)$ as a function of temperature $T$ for different $\varphi$. (d) $E_{m_2|b_1}(\tau)$ in the coupling strength $g$ and decay rate $\kappa$ parameter space. (e) $E_{m_2|b_1}(\tau)$ in the parameter space spanned by magnon detunings $\Delta_1$ and $\Delta_2$. For (b)-(e), the evolution time is set to $\tau=3~\mu$s. The temperature is $T=10$ mK, except for panel (c). $g/2\pi=2$ MHz for (c) and (e). $\kappa/2\pi=3$ MHz except for (d). $\varphi=\pi$ for (d) and (e). $\Delta_1/2\pi=\Delta_2/2\pi=-10$ MHz for (a)-(d). Other parameters are set as $\omega_b/2\pi=10$ MHz, $\epsilon/2\pi=10$ GHz, $\kappa_b/2\pi=100$ Hz, $\gamma/2\pi=1$ MHz, and $g_2=0$.}\label{MPent}
\end{figure}
We first consider the process of remotely preparing entanglement between a single magnon and a single phonon, corresponding to a system with $N=2$ YIG spheres. The parameters are chosen as $\Delta_1=\Delta_2=-\omega_b$, $\varphi_{12}=\varphi$, $\kappa_1=\kappa_2=\kappa$, $g_1=g$, and $g_2=0$. Under these conditions, in the rotating frame defined by the free Hamiltonian, applying the rotating-wave approximation to Eq.~\eqref{Hamsystem} yields the effective Hamiltonian
\begin{equation}
	\label{Heff}
	H_{\rm eff}=g(m^\dag_1b^\dag_1+mb)-i\kappa e^{i\varphi}(m^\dag_1m_2+m_1m^\dag_2).
\end{equation}
The parametric down-conversion process, corresponding to the term $m_1^\dag b_1^\dag+\mathrm{h.c.}$, generates quantum entanglement between $m_1$ and $b_1$. Meanwhile, the beam-split interaction, i.e., $m_1^\dag m_2+\mathrm{h.c.}$, enables the transfer of this entanglement from the subsystem $m_1-b_1$ to the remote bipartite subsystem $m_2-b_1$. In addition, owing to the periodicity of the trigonometric functions entering the magnon-magnon coupling, integer multiples of $2\pi$ in the propagation phase are neglected throughout the discussion, i.e., $2k\pi+\varphi\equiv\varphi$.

Based on the CM dynamics in Eq.~\eqref{cmvt} and the definition of LN in Eq.~\eqref{lognegayivity}, the two-mode entanglement can be evaluated numerically, with the results presented in Fig.~\ref{MPent}. In Fig.~\ref{MPent}(a), the dynamical generation process of remote bipartite magnon-phonon entanglement, denoted by $E_{m_2|b_1}(t)$, is illustrated for various values of the magnomechanical coupling strength $g$ and traveling phase $\varphi$. It is observed that, irrespective of the values of $g$ and $\varphi$, the LN $E_{m_2|b_1}$ gradually converges to a steady value over time, signifying the establishment of stable entanglement within the two modes $m_2$ and $b_1$. Furthermore, for $\varphi=\pi/2$ (blue solid line), which corresponds to purely coherent coupling, the steady-state entanglement increases noticeably with increasing $g$, taking values of $0.077$, $0.099$, and $0.114$, respectively. In contrast, for $\varphi=\pi$ (purple dash-dotted line), corresponding to purely dissipative coupling, the LNs are $0.229$, $0.229$, and $0.221$, respectively, showing only negligible variation with $g$. Notably, the entanglement for $\varphi=\pi$ remains consistently stronger than that for $\varphi=\pi/2$ across all values of $g$.

To examine the impact of parameters on entanglement generation, the LN $E_{m_2|b_1}(\tau)$ at $\tau=3~\mu$s is shown in Figs.~\ref{MPent}(b)-(e). Figure~\ref{MPent}(b) shows the dependence of the LN $E_{m_2|b_1}(\tau)$ on the coupling strength $g$ at temperature $T=10$ mK. For a given $g$, $E_{m_2|b_1}(\tau)$ attains its maximum at $\varphi=\pi$, and exceeds the values obtained at $\varphi=3\pi/4$ (mixed dissipative and coherent coupling) and $\varphi=\pi/2$. Additionally, for $\varphi=\pi/2$, $E_{m_2|b_1}(\tau)$ increases monotonically with $g$, whereas for $\varphi=\pi$, an optimal coupling range of $1.9~\text{MHz}\leq g/2\pi\leq 2.3$ MHz yields enhanced entanglement. The temperature dependence of $E_{m_2|b_1}(\tau)$ at fixed coupling strength $g/2\pi=2$ MHz is presented in Fig.~\ref{MPent}(c). A similar trend is observed. The LN is maximized at $\varphi=\pi$, takes intermediate values at $\varphi=3\pi/4$, and reaches its minimum at $\varphi=\pi/2$. Notably, the system at $\varphi=\pi$ exhibits more resilience to thermal noise, with $E_{m_2|b_1}(\tau)\approx0.045$ even at $T=200$ mK, while the LNs for $\varphi=\pi/2$ and $\varphi=3\pi/4$ vanish at lower temperatures. Figure~\ref{MPent}(d) shows the LN $E_{m_2|b_1}(\tau)$ versus the coupling strength $g$ and magnon decay rate $\kappa$. The entanglement increases approximately with both parameters and exceeds $0.25$ for $g/2\pi\ge2$ MHz and $\kappa/2\pi\ge3.5$ MHz. To assess the robustness of the proposed scheme against detuning imperfections, we further analyze the entanglement behavior under finite magnon detunings, as shown in Fig.~\ref{MPent}(e). Specifically, Fig.~\ref{MPent}(e) displays $E_{m_2|b_1}(\tau)$ in the parameter space spanned by $\Delta_1$ and $\Delta_2$, revealing an optimal region for $-1.2\omega_b\le\Delta_{1,2}\le-0.8\omega_b$, with a slightly broader tolerance for $\Delta_1$ than for $\Delta_2$.

\subsection{Multimode entanglement}\label{Multimode}
\begin{figure}[t]
	\centering
	\includegraphics[width=0.48\textwidth]{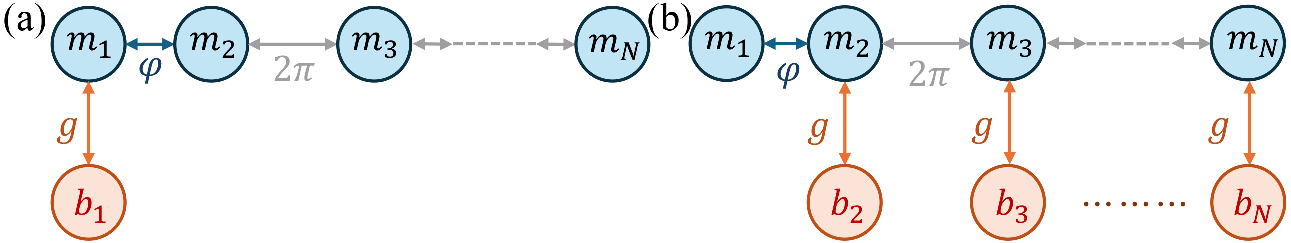}
	\includegraphics[width=0.235\textwidth]{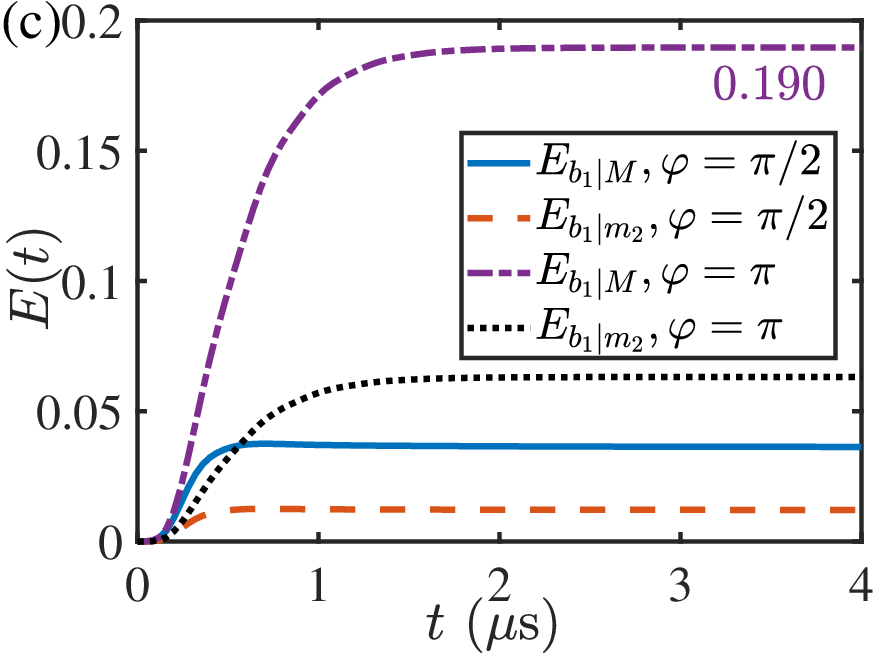}
	\includegraphics[width=0.235\textwidth]{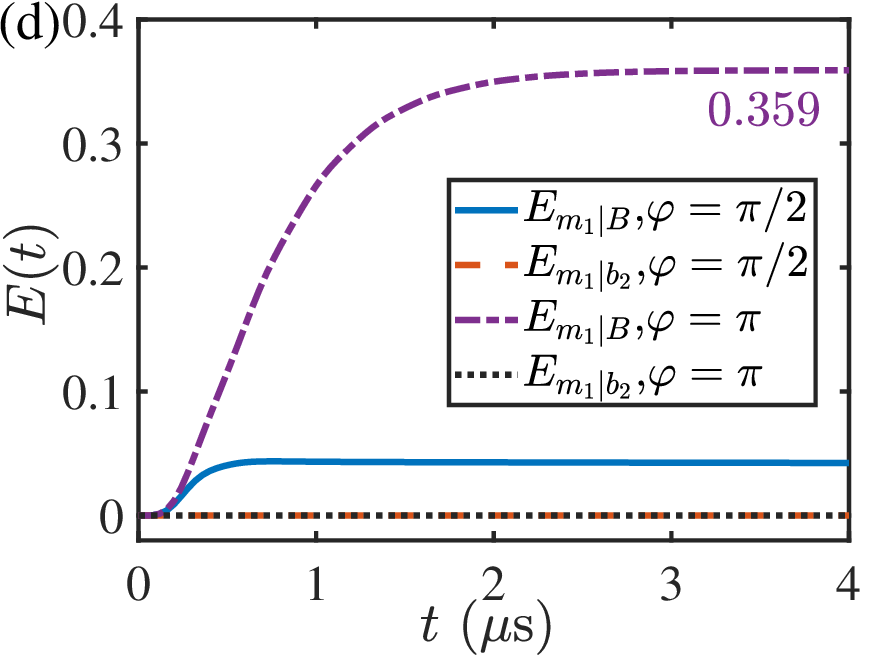}
	\includegraphics[width=0.235\textwidth]{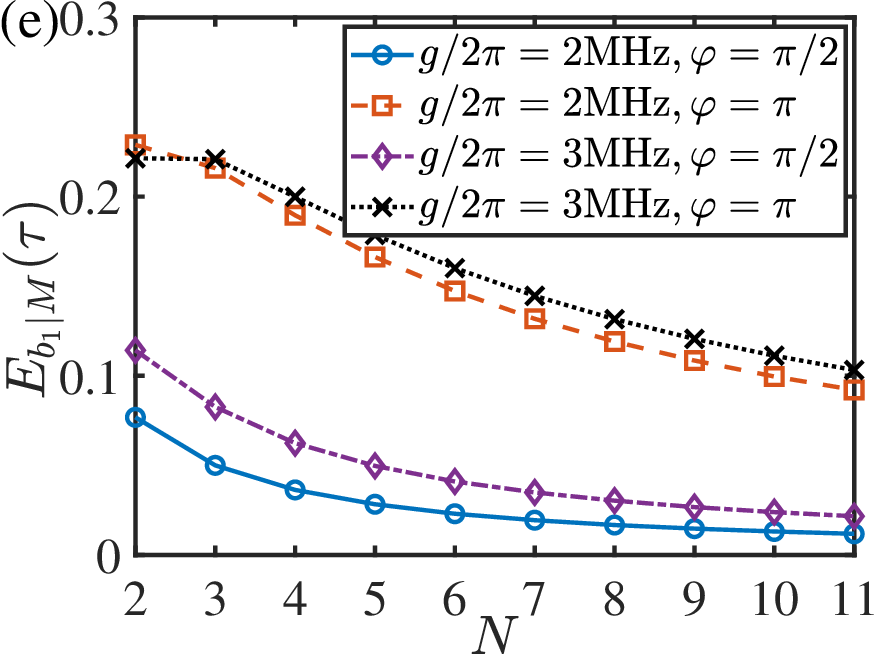}
	\includegraphics[width=0.235\textwidth]{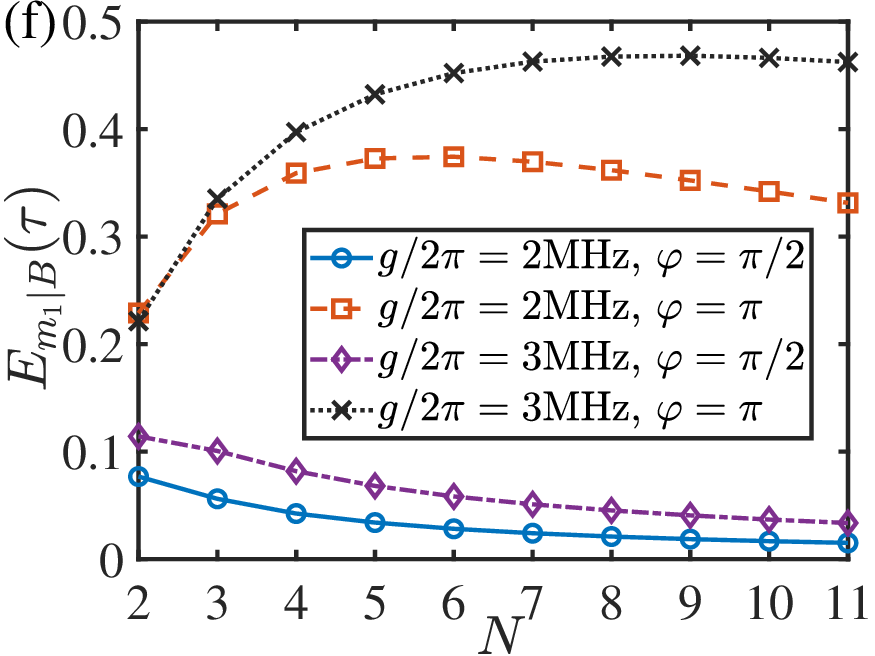}
	\caption{[(a), (b)] Schematic diagrams illustrating the entanglement between a single phonon mode and $N$ magnon modes, and between a single magnon mode and $N$ phonon modes, respectively. [(c), (d)] Time evolution of the LNs $E(t)$ under different phases for the models shown in (a) and (b), respectively. [(e), (f)] Dependence of the LNs $E_{b_1|M}(\tau)$ and $E_{m_1|B}(\tau)$ on the number of modes $N$, corresponding to the models in (a) and (b), respectively. Here, $g/2\pi=2~\mathrm{MHz}$ and $N=4$ for (c) and (d), while $\tau=4~\mu\mathrm{s}$ for (e) and (f). All remaining parameters are identical to those used in Fig.~\ref{MPent}(a).}\label{Multient}
\end{figure}

In the preceding section, we demonstrated the generation of remote entanglement between a single magnon mode and a single phonon mode. We now turn our attention to the feasibility of generating remote multipartite entangled states involving magnons and phonons. For simplicity and without loss of generality, all magnon modes are assumed to have identical detunings and identical decay rates, i.e.,
$\Delta_1=\Delta_2=\cdots=\Delta_N=\Delta$ and $\kappa_1=\kappa_2=\cdots \kappa_N=\kappa$. 

Specifically, we first consider the entanglement between a single phonon mode and $N$ magnon modes. As shown in Fig.~\ref{Multient}(a), the parameters in Fig.~\ref{diagram} are chosen as $g_1=g$, $g_2=g_3=\cdots=g_N=0$, $\varphi_{12}=\varphi$, and $\varphi_{23}=\varphi_{34}=\cdots=\varphi_{(N-1)N}=2\pi$. The corresponding system Hamiltonian and drift matrix are provided in Eqs.~\eqref{Hamapho} and~\eqref{Adriftapho} of Appendix~\ref{appc}, respectively. Numerical results under blue-detuning regime $\Delta=-\omega_b$ are shown in Figs.~\ref{Multient}(c) and (e). In Fig.~\ref{Multient}(c), we present the dynamical evolution of the LNs for different phases $\varphi$ at a fixed $N=4$. Here, $E_{b_1|M}\equiv E_{b_1|m_2\cdots m_N}$ quantifies the bipartite entanglement between the phonon mode $b_1$ and all magnon modes except $m_1$. As the system evolves, all LNs approach their respective steady values asymptotically. Specifically, for purely dissipative coupling between $m_1$ and $m_2$ ($\varphi=\pi$), the LNs $E_{b_1|M}$ and $E_{b_1|m_2}$ converge to approximately $0.190$ and $0.063$, respectively. By contrast, under purely coherent coupling ($\varphi=\pi/2$), these values are reduced to $0.036$ and $0.012$, respectively. This result further supports the conclusion that dissipative interactions enhance entanglement generation. Figure~\ref{Multient}(e) illustrates the dependence of $E_{b_1|M}$ at time $\tau=4~\mu$s on the number of magnon modes $N$. For a fixed $N$, the LN $E_{b_1|M}$ increases with the magnomechanical coupling strength $g$, irrespective of whether the interaction between $m_1$ and $m_2$ is coherent or dissipative. In addition, $E_{b_1|M}$ decreases monotonically with increasing $N$. In the coherent-coupling regime ($\varphi=\pi/2$), the rate of decrease gradually diminishes and eventually saturates, whereas in the dissipative-coupling regime ($\varphi=\pi$), the decline remains nearly linear with an approximately constant slope. For instance, at $g/2\pi=3$ MHz, $E_{b_1|M}$ decreases from $0.200$ ($N=4$) to $0.103$ ($N=11$) for $\varphi=\pi$, while it is reduced from $0.062$ to $0.022$ for $\varphi=\pi/2$. Moreover, due to the prescribed phase relations, one can demonstrate that $E_{b_1|m_2}=E_{b_1|m_3}=\cdots =E_{b_1|m_N}$ and $E_{b_1|M}=(N-1)E_{b_1|m_2}$, which numerical results can be found in Fig.~\ref{appcfig} of Appendix~\ref{appc}. Moreover, since $E_{b_1|m_j}>0$ for any $j=2,\cdots N$, it follows that no bipartition of the hybrid system is separable, indicating the presence of genuine remote multimode entanglement.

Furthermore, we investigate the entanglement between a single magnon mode and $N$ phonon modes. As illustrated in Fig.~\ref{Multient}(b), the parameters in Fig.~\ref{diagram} are chosen as $g_1=0$, $g_2=g_3=\cdots g_N=g$, $\varphi_{12}=\varphi$, and $\varphi_{23}=\varphi_{34}=\cdots=\varphi_{(N-1)N}=2\pi$. The corresponding system Hamiltonian and drift matrix are given in Eqs.~\eqref{Hamamag} and~\eqref{Adriftamag} of Appendix~\ref{appc}. Numerical results under $\Delta=-\omega_b$ are presented in Figs.~\ref{Multient}(d) and (f). The time evolutions of the LNs $E_{m_1|B}$ and $E_{m_1|b_2}$ at $N=4$ are shown in Fig.~\ref{Multient}(d), where $E_{m_1|B}\equiv E_{m_1|b_2\cdots b_N}$ denotes the LN between the magnon mode $m_1$ and all phonon modes. Similar to the behavior in Fig.~\ref{Multient}(c), $E_{m_1|B}$ approaches a steady value in the long time limit, reaching $0.359$ at $\varphi=\pi$, which significantly exceeds the value $0.043$ obtained at $\varphi=\pi/2$. In contrast, the LN $E_{m_1|b_2}$ remains zero throughout the entire evolution. The value of $E_{m_1|B}$ at $\tau=4~\mu$s is plotted in Fig.~\ref{Multient}(f). In the coherent-coupling regime ($\varphi=\pi/2$), $E_{m_1|B}$ decreases monotonically with increasing $N$, with the decreasing rate gradually slowing and eventually saturating. In the dissipative-coupling regime ($\varphi=\pi$), $E_{m_1|B}$ initially increases and then decreases as $N$ grows. For $g/2\pi=2$~MHz, the optimal range of magnon numbers lies in the range $4\le N\le7$, with the LN reaching a maximum value of $0.374$ at $N=5$. For $g/2\pi=3$~MHz, this optimal range shifts to $7\le N\le 10$, with the maximum $0.468$ attained at $N=8$. Thus, a stronger coupling $g$ allows a larger ensemble of magnomechanical subsystems to participate in the enhancement of entanglement. Additional numerical results for other bipartitions are provided in Appendix~\ref{appc}.

\subsection{Genuine four-mode entanglement}
\begin{figure}[t]
	\centering
	\includegraphics[width=0.235\textwidth]{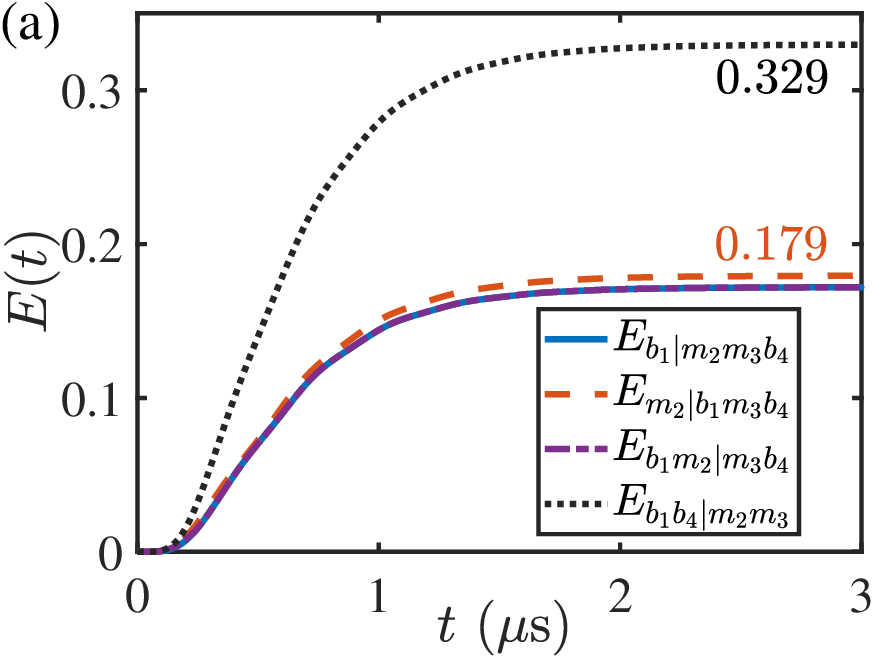}
	\includegraphics[width=0.235\textwidth]{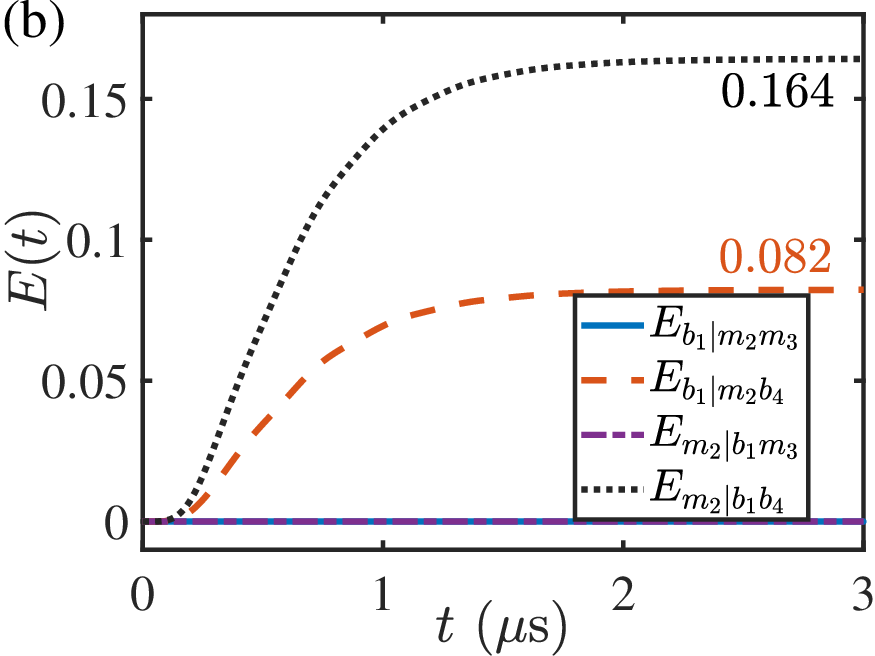}
	\includegraphics[width=0.235\textwidth]{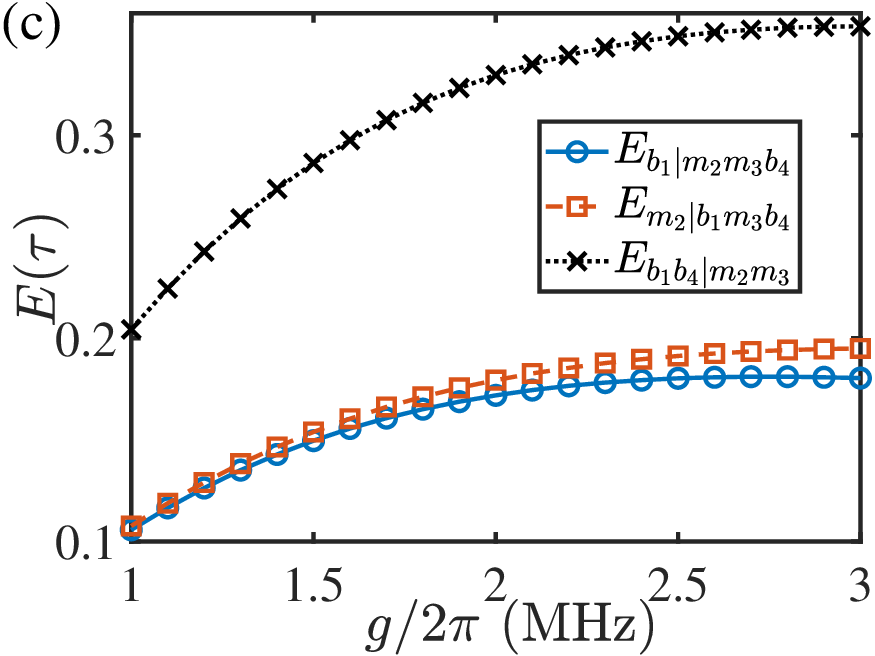}
	\includegraphics[width=0.235\textwidth]{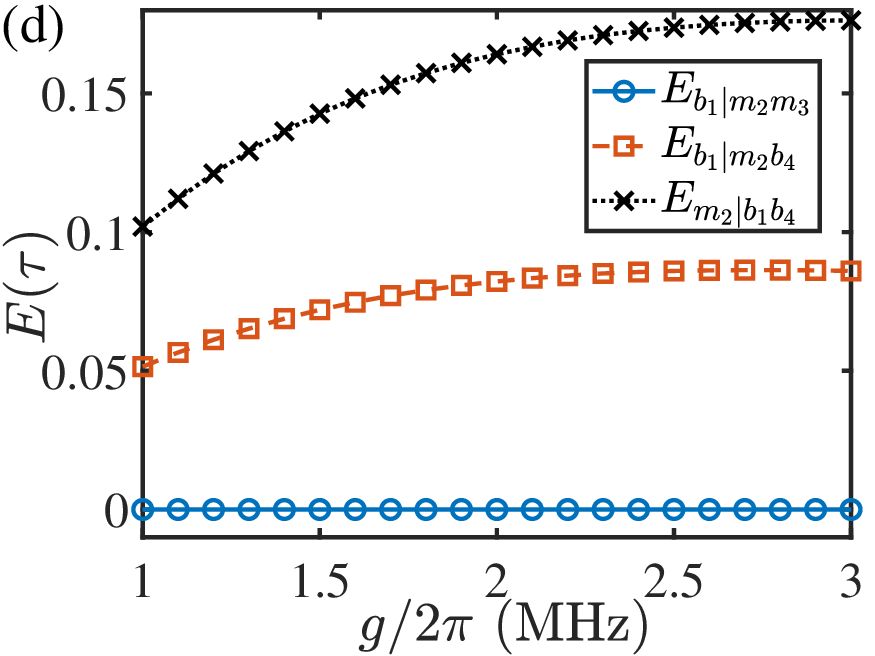}
	\caption{[(a), (b)] The time evolution of the LNs $E(t)$ associated with the genuine four-mode entanglement is shown, with $g/2\pi=2$ MHz. [(c), (d)] Dependence of the various LNs $E(\tau)$ at fixed time $\tau=3~\mu$s on the coupling strength $g$. Here, the remaining parameters are identical to those in Fig.~\ref{MPent}(a).}\label{fourmp}
\end{figure}

In this section, we consider a hybrid model consisting of four magnon modes and two phonon modes, where the parameters shown in Fig.~\ref{diagram} are fixed as $N=4$, $g_1=g_4=g$, $g_2=g_3=0$, and $\varphi_{12}=\varphi_{23}=\varphi_{34}=2\pi$. The system Hamiltonian in Eq.~\eqref{Hamsystem} under this case can be simplified as
\begin{equation}\label{Hamfour}
\begin{aligned}
H&=\sum^4_{j=1}\Delta m^\dag_jm_j+\sum_{j=1,4}\omega_bb^\dag_jb_j+g(m_j+m^\dag_j)(b_j+b^\dag_j)\\
&\quad-i\kappa\sum_{j=1}^3 \sum^4_{l=j+1}(m^\dag_j m_l+m_jm^\dag_l).
\end{aligned}
\end{equation}
Here, we assume $\Delta_1=\Delta_2=\Delta_3=\Delta_4=\Delta$ and $\kappa_1=\kappa_2=\kappa_3=\kappa_4=\kappa$. The corresponding drift matrix A can be obtained from Eq.~\eqref{driftANmag}, after which the CM $V(t)$ is calculated according to Eq.~\eqref{cmvt}. Based on $V(t)$, the LNs characterizing the hybrid system are obtained, and the corresponding numerical results are shown in Fig.~\ref{fourmp}. The detuning is set to $\Delta=-\omega_b$.

In Fig.~\ref{fourmp}(a), we plot the time evolution of the LNs $E_{b_1|m_2m_3b_4}$ (blue solid line), $E_{m_2|b_1m_3b_4}$ (red dashed line), $E_{b_1m_2|m_3b_4}$ (purple dash-dotted line), and $E_{b_1b_4|m_2m_3}$ (black dotted line) for the entire four-mode system. It is evident that all LNs converge to nonzero constants after sufficiently long evolution. In particular, $E_{b_1b_4|m_2m_3}$ attains the largest value, reaching approximately $0.329$, while $E_{m_2|b_1m_3b_4}$ approaches to $0.179$, slightly exceeding the remaining LNs $E_{b_1|m_2m_3b_4}$ and $E_{b_1m_2|m_3b_4}$. According to the symmetry relations given in Eq.~\eqref{Hamfour}, one obtains $E_{b_4|b_1m_2m_3}=E_{b_1|m_2m_3b_4}$, $E_{m_3|b_1m_2b_4}=E_{m_2|b_1m_3b_4}$, and $E_{b_1m_2|m_3b_4}=E_{b_1m_3|m_2b_4}$. These results confirm that every possible bipartition of the four-mode system is entangled, evidencing the existence of genuine four-mode entanglement. The values of the LNs at $\tau=3~\mu$s are plotted in Fig.~\ref{fourmp}(c). The LN $E_{b_1m_2|m_3b_4}$ is not shown, as it is always equal to $E_{b_1|m_2m_3b_4}$ for all parameter settings. The results indicate that the LNs approximately gradually increase as $g$ is enhanced, although the rate of increase progressively slows for larger $g$. Notably, for $g/2\pi>2.8$ MHz, $E_{b_1|m_2m_3b_4}$ begins to decrease slightly, suggesting that further increases in $g$ may not always enhance this LN and can instead reduce its value. Furthermore, the LNs satisfy the relation $E_{b_1|m_2m_3b_4}<E_{m_2|b_1m_3b_4}<E_{b_1b_4|m_2m_3}$, indicating that the bipartite entanglement between the two phonon modes and the two magnon modes is the strongest. 

Moreover, we illustrate the bipartite entanglement within three-mode subsystems in Fig.~\ref{fourmp}(b), corresponding to the total system Hamiltonian in Eq.~\eqref{Hamfour}, to characterize entanglement between different parts of the system. As the system evolves, stable entanglement progressively develops between the phonon $b_1$ and the magnon-phonon $m_2b_4$, as well as between $m_2$ and the biphonon $b_1b_4$. Specifically, it is observed that $E_{b_1|m_2b_4}<E_{m_2|b_1b_4}$, with corresponding values of $0.082$ and $0.164$, respectively. Conversely, no entanglement is generated between $b_1$ and $m_2m_3$, nor between $m_2$ and $b_1m_3$. More LNs can be similarly obtained by the symmetry relations, such as $E_{b_1|m_2b_4}=E_{b_1|m_3b_4}=E_{b_4|m_2b_1}=E_{b_4|m_3b_1}$. The corresponding values of $E(\tau)$ are similarly shown in Fig.~\ref{fourmp}(d). One can observe that both $E_{b_1|m_2b_4}$ and $E_{m_2|b_1b_4}$ increase as the $g$ is enhanced, and the increment diminishes progressively as $g$ continues to increase. The values satisfy $E_{b_1|m_2b_4}<E_{m_2|b_1b_4}$ and $E_{b_1|m_2m_3}=0$ for all coupling strengths. Furthermore, we also numerically demonstrate that $E_{b_1|m_2}=E_{b_1|b_4}=E_{m_2|m_3}=0$, which indicates that an arbitrary bipartition between one-mode and one-mode is separable. These results reveal that the phonon modes $b_1$ and $b_4$ behave as a collective subsystem capable of establishing entanglement with each magnon mode. The presence of the other phonon mode prevents the first phonon mode from entangling with the distinct magnon modes.

\section{Discussion and Conclusion}\label{Secconclu}
Our protocol is based on a hybrid waveguide magnomechanical system that combines waveguide photon-magnon interactions~\cite{Anomagnon,giantspin,SingleMode,Unidirectional} with magnomechanical coupling~\cite{magnoncavity,kerrcavitymagnon,cavitymagnomechan}. All parameters used here are compatible with recent experiments. The magnon mode possesses a transition frequency of about $10$~GHz, with an intrinsic decay rate of $\gamma\sim 1$~MHz and a radiative decay rate $\kappa$ on the order of several MHz~\cite{giantspin,SingleMode,Unidirectional}. For the phonon mode, the characteristic frequency is around $10$~MHz, accompanied by a much smaller decay rate of $\kappa_b\sim 100$~Hz~\cite{magnoncavity,kerrcavitymagnon}. The single-excitation magnon-phonon coupling $\tilde{g}$ depends on the volume of the YIG sphere and the orientation of the bias magnetic field. For a YIG sphere with a diameter of $100~\mu$m, the coupling strength is on the order of $0.1$ Hz~\cite{magnoncavity}. Under strong driving, the Rabi frequency can reach $\Omega\sim 10^{14}$ Hz, yielding a steady-state magnon amplitude of $\langle m\rangle \approx 10^7$~\cite{mppentang,magnonkerreff}. Consequently, the driving-enhanced magnomechanical coupling can be increased to $g=\tilde{g}\langle m\rangle \sim 1$~MHz~\cite{mppentang}. The detuning $\Delta$ is governed by the external drive and can be readily adjusted via the driving frequency. In particular, the required blue-detuned regime has been experimentally demonstrated~\cite{magnoncavity,kerrcavitymagnon}. In addition, the cryogenic environment has been estimated in recent studies~\cite{polaritonphonon}. Moreover, all parameters are assumed to be independent of the environmental temperature in the simulations, and this choice is consistent with previous works~\cite{mppentang}.

Moreover, recent studies show that a single magnon mode can couple to a microstrip meandering waveguide at two spatially separated points~\cite{giantspin}, enabling tunable radiative decay via the coupling distance. This provides additional control over entanglement and suggests promising directions for future work. Superconducting resonators have also been experimentally coupled to magnon modes in YIG spheres~\cite{coherremag} and interfaced with YIG films~\cite{YIGfilm}, indicating that superconducting coplanar waveguides offer a viable platform for implementing our protocol. Lastly, we discuss the detection and verification of entanglement. Following established methods~\cite{entdetect,entdeteopto}, the generated entanglement can be characterized by reconstructing the corresponding CM. The magnon state can be probed with a weak microwave field, while the mechanical quadratures can be measured via an auxiliary optical cavity driven by a weak red-detuned field~\cite{mppentang}, enabling verification of the magnon-phonon entanglement. Furthermore, the scheme can be extended to waveguide optomechanical systems, where multiple cavities coupled to a common waveguide enable remote photon-phonon entanglement.

In summary, we have proposed a protocol for generating remote magnon-phonon entanglement in a waveguide magnomechanical system, where magnons in spatially separated YIG spheres are coupled both to a common waveguide and to their local phonon modes. By appropriately engineering the driving-enhanced magnomechanical interactions, the scheme enables the generation of diverse forms of remote and dynamically stable entanglement. These include fundamental two-mode magnon-phonon entanglement, as well as more complex configurations such as genuine multimode entanglement between a single phonon and multiple magnons, bipartite entanglement between a single magnon and multiple phonons, and genuine four-mode entanglement involving two magnons and two phonons. Moreover, our analysis shows that waveguide-photon-mediated dissipative magnon-magnon interactions generate substantially stronger entanglement than coherent coupling. Our work presents an interesting and experimentally feasible framework for generating remote entanglement in hybrid quantum systems.

\section*{Acknowledgments}
Shi-fan Qi and Fan Li world like to thank Yan-Kui Bai for helpful discussions. We acknowledge financial support from the National Science Foundation of China (Grant No. 12404405), Hebei National Science Foundation (Grant No. A2025205030), the Science Research Project of Hebei Education Department (Grant No. NJ2026194), Hebei 333 Talent Project (Grant No. B20231005), and the funds of Hebei Normal University (Grant No. L2024B10 and No. L2026J02).

\section*{Data availability}
The data support the findings of this article are not publicly available. The data are available from the authors upon reasonable request.

\begin{widetext}
\appendix
\section{The magnon-magnon coupling mediated by waveguide photons}\label{appa}
In this appendix, we derive the magnon-magnon interactions mediated by traveling photons. The waveguide magnon system under consideration consists of $N$ YIG spheres positioned at distinct positions along a one-dimensional microstrip waveguide~\cite{giantspin,Anomagnon,SingleMode,Unidirectional}. The magnon modes provided by the YIG spheres couple with the traveling photons through magnetic dipole interaction. The total Hamiltonian of this system is given by
\begin{equation}\label{Hmfree}
\begin{aligned}
	H=\sum^N_{j=1}\omega_jm^\dag_jm_j+\int^{+\infty}_{-\infty}dk\tilde{\omega}_ka^\dag_ka_k+\sum^N_{j=1}\int^{+\infty}_{-\infty}dk(G_je^{-ikL_j}m_ja^\dag_k+{\rm h.c.}),
\end{aligned}
\end{equation}
where $a_k$ ($a^\dag_k$) is the annihilation (creation) operator of photons with transition frequency $\omega_k$ and wave-vector $k$. $G_j$ is the coupling strength between magnon-mode $m_j$ and photon modes. $L_j$ is the distance between the coupling points $m_1$ and $m_j$, and $L_1=0$.  With respect to the transformation $U=\exp\{i\int^{+\infty}_{-\infty}dk\tilde{\omega}_ka^\dag_ka_kt\}$, the original Hamiltonian in Eq.~\eqref{Hmfree} turns into 
\begin{equation}\label{Hmint}
	H_{\rm int}=\sum^N_{j=1}\omega_jm^\dag_jm_j+\sum^N_{j=1}\int^{+\infty}_{-\infty}dk(G_je^{-ikL_j}e^{i\tilde{\omega}_kt}m_ja^\dag_k+{\rm h.c.}).
\end{equation}

To study the coherent and dissipative dynamics of the multi-magnon subsystem, within the framework of the Markovian approximation, we eliminate the waveguide field using a standard procedure~\cite{openquantum,atomfield} and derive the following master equation for the reduced density operator
\begin{equation}\label{rhosolu}
	\partial_t\rho(t)=-\int^\infty_0 d\tau {\rm Tr}_w\left[H_{\rm int}(t),[H_{\rm int}(t-\tau), \rho_w\otimes\rho(t)]\right],
\end{equation}
where ${\rm Tr}_w$ represents a partial tracing over the waveguide degrees of freedom, and $\rho_w=|0\rangle \langle 0|$ denotes the initial vacuum state of the waveguide's photonic modes. $\rho(t)$ is the density matrix of the multi-magnon modes. By substituting the interaction Hamiltonian $H_{\rm int}$ into Eq.~\eqref{rhosolu}, we obtain
\begin{equation}\label{rhodot}
\begin{aligned}
	\partial_t\rho(t)&=-i\sum^N_{j=1}\left[\omega_jm^\dag_jm_j,\rho(t)\right]-i\sum^{N-1}_{j=1}\sum^N_{l=j+1}\sqrt{\kappa_j\kappa_l}\sin\varphi_{jl}\left[m^\dag_jm_l+m_jm^\dag_l,\rho(t)\right]\\
	&+\sum^N_{j=1}\kappa_j[2m_j\rho(t) m^\dag_j-m^\dag_jm_j\rho(t)-\rho(t) m^\dag_jm_j]+\sum^{N-1}_{j=1}\sum^N_{l=j+1}\sqrt{\kappa_j\kappa_l}\cos\varphi_{jl}[2m_j\rho(t) m_l^\dag-m^\dag_jm_l\rho(t)-\rho(t)m^\dag_jm_l],\\
\end{aligned}
\end{equation}
with
\begin{equation}\label{gamma}
	\kappa_j=G_j^2\int^\infty_0 d\tau\int^{+\infty}_{-\infty} dk e^{\pm i\tilde{\omega}_k\tau}=2\pi G^2_j/v_g,
\end{equation}
by $\int^{+\infty}_{-\infty} e^{\pm ikx}dx=2\pi\delta(x)$. $\kappa_j$ represents the radiative decay rate associated with waveguide photons. The phase difference is defined as $\varphi_{jl}=\varphi_l-\varphi_j$, where $\varphi_j$ denotes the traveling phase, with $\varphi_1=0$ and $\varphi_j=\omega_j L_j/v_g$ for $j=2,3,\cdots,N$. Here, $v_g$ represents the group velocity of the microwave field. Under the condition that all magnon modes share a common transition frequency $\omega_m$, the phase difference simplifies to $\varphi_{jl}=\omega_mL_{jl}/v_g$, where $L_{jl}$ denotes the distance between two YIG spheres. This relation remains valid when $\omega_j \approx\omega_l\approx\omega_m$, as demonstrated in recent experiments~\cite{giantspin}. From a physical perspective, a microwave photon emitted by one magnon must propagate a finite distance through the waveguide before being absorbed by another magnon mode, thereby acquiring an additional phase. This process induces a propagation-dependent phase in the magnon-magnon coupling mediated by these traveling photons.

From the master equation in Eq.~\eqref{rhodot}, it is evident that the coupling between magnon modes $m_j$ and $m_l$, mediated by the waveguide photons, can be divided into two components, the dissipative coupling $\sqrt{\kappa_j\kappa_l}\cos\varphi_{jl}$ and the coherent coupling $\sqrt{\kappa_j\kappa_l}\sin\varphi_{jl}$. The equation~\eqref{rhodot} corresponds to a non-Hermitian Hamiltonian
\begin{equation}
	H_{m}=\sum^N_{j=1}\omega_jm^\dag_jm_j-i\sum^{N-1}_{j=1}\sum^N_{l=j+1}\sqrt{\kappa_j\kappa_l}e^{i\varphi_{jl}}(m^\dag_jm_l+m_jm_l^\dag).
\end{equation}

\section{The derivation of the linearized Hamiltonian}\label{appb}
This appendix presents the derivation of the system linearized Hamiltonian in Eq.~\eqref{Hamsystem}. Under the rotating frame with $U=\exp[i\epsilon t\sum^N_{j=1}m^\dag_jm_j]$, the system Hamiltonian in Eq.~\eqref{Htotorigin} turns into
\begin{equation}\label{Htotinter}
\begin{aligned}
H'_{\rm tot}=\sum^N_{j=1}\Delta_jm^\dag_jm_j+\omega_bb^\dag_jb_j+\tilde{g}_jm^\dag_jm_j(b_j+b^\dag_j)-i\sum^{N-1}_{j=1}\sum^N_{l=j+1}\sqrt{\kappa_j\kappa_l} e^{i\varphi_{jl}}(m^\dag_jm_l+m_jm^\dag_l)+\sum^N_{j=1}\Omega_j(m^\dag_j+m_j),
\end{aligned}
\end{equation}
where $\Delta_j=\omega_j-\epsilon$ is the detuning of magnon $m_j$. Using the standard QLE and the master equation in Eq.~\eqref{rhodot}, the time evolution of the system operators under the Hamiltonian in Eq.~\eqref{Htotorigin} is given by
\begin{equation}\label{operdot}
\begin{aligned}
	\dot{m}_j&=-(i\Delta_j+\tilde{\kappa}_j)m_j-i\tilde{g}_jm_j(b_j+b^\dag_j)-\sum^N_{l\neq j}(\sqrt{\kappa_j\kappa_l}e^{i\varphi_{jl}}m_l-\sqrt{2\Gamma_{jl}}m_{l_{\rm in}})
	+\sqrt{2\tilde{\kappa}_j}m_{j_{\rm in}}-i\Omega_j,\\
	\dot{b}_j&=-(i\omega_b+\kappa_{b_j})b_j-i\tilde{g}_jm^\dag_j m_j+\sqrt{2\kappa_{b_j}}b_{j_{\rm in}},\\
\end{aligned}
\end{equation}
where $\tilde{\kappa_j}=\kappa_j+\gamma$ is the total decay rate of mode $m_j$, with $\gamma$ denoting the intrinsic decay rate. $\Gamma_{jl}\equiv\sqrt{\kappa_j\kappa_l}\cos\varphi_{jl}$ is the dissipative coupling strength between magnon modes $m_j$ and $m_l$. $m_{j_{\rm in}}$ and $b_{j_{\rm in}}$ are the input noise operators for $m_j$ and $b_j$, respectively, which are characterized by their corresponding functions: $\langle o^\dag_{\rm in}(t)o_{\rm in}(t')\rangle=\bar{n}_{o}\delta(t-t')$ and $\langle o_{\rm in}(t)o^\dag_{\rm in}(t')\rangle=[\bar{n}_{o}+1]\delta(t-t')$. $\bar{n}_{o}=[{\rm exp}(\hbar\omega_o/k_B T)-1]^{-1}$ is the mean population number at the thermal equilibrium state, with $o=m_j,b_j$. For simplicity, we assume $\bar{n}_{m_j}=\bar{n}$, $\kappa_{b_j}=\kappa_b$, and $\bar{n}_{b_j}=\bar{n}_b$ in what follows and throughout the main text.

Under the condition of strong driving pulse $\Omega_j$, the magnon mode $m_j$ has a large steady-state amplitude, i.e., $|\langle m_j\rangle|\gg 1$. This allows us to linearize the system's dynamics around the constant values by expressing the operators as $o=\langle o\rangle+\delta o$, where $o=m_j,b_j$ and $\delta o$ represents the operator describing the quantum fluctuation. The constant values $\langle o\rangle$ satisfy
\begin{equation}\label{steadyvalue}
\begin{aligned}
	(i\Delta_j+\tilde{\kappa}_j)\langle m_j\rangle+i\tilde{g}_j\langle m_j\rangle(\langle b_j\rangle+\langle b_j\rangle^*)+\sum^N_{l\neq j}i\sqrt{\kappa_j\kappa_l}e^{i\varphi_{jl}}\langle m_l\rangle+i\Omega_j&=0,\\
	(i\omega_b+\kappa_b)\langle b_j\rangle+i\tilde{g}_j|\langle m_j\rangle|^2&=0.\\
\end{aligned}
\end{equation}
When $\Delta_j\gg \tilde{\kappa}_j,\tilde{g}_j$, the expectation value $\langle m_j\rangle$ is approximately given by $\langle m_j\rangle\approx-\Omega_j/\Delta_j$, which is purely real.

By substituting the constant values in Eq.~\eqref{steadyvalue} into the equations in Eq.~\eqref{operdot} and ignoring all the higher-order terms of fluctuations, the QLEs describing the fluctuation operator $\delta o$ can be described as
\begin{equation}\label{deltaoper}
\begin{aligned}
	\dot{\delta m}_j&=-(i\tilde{\Delta}_j+\tilde{\kappa}_j)\delta m_j-i\tilde{g}_j\langle m_j\rangle(\delta b_j+\delta b^\dag_j)-\sum^N_{l\neq j}(\sqrt{\kappa_j\kappa_l}e^{i\varphi_{jl}}\delta m_l-\sqrt{2\Gamma_{jl}}m_{l_{\rm in}})+\sqrt{2\tilde{\kappa}_j}m_{j_{\rm in}},\\
	\dot{\delta b}_j&=-(i\omega_b+\kappa_b)\delta b_j-i\tilde{g}_j(\langle m_j\rangle^*\delta m_j+\langle m_j\rangle\delta m^\dag_j)+\sqrt{2\kappa_b}b_{j_{\rm in}},
\end{aligned}
\end{equation}
where $\tilde{\Delta}_j=\Delta_j+\tilde{g}_j(\langle b_j\rangle+\langle b_j\rangle^*)$ is the effective detuning. It should be noted that the nonlinear coupling term $\tilde{g}_j\delta m_j\delta m^\dagger_j(\delta b_j+\delta b_j^\dagger)$ is neglected, as its coupling strength $\tilde{g}_j$ is significantly smaller than the driving-enhanced linear coupling strength $\tilde{g}_j\langle m_j\rangle$. Under strong driving, $\langle m_j\rangle\sim 10^7$, leading to a difference of approximately seven orders of magnitude. Accordingly, the nonlinear interaction terms can be safely neglected. The corresponding effective linearized Hamiltonian can be described as
\begin{equation}
\begin{aligned}
	H_{\rm lin}&=\sum^N_{j=1}\tilde{\Delta}_j\delta m^\dag_j\delta m_j+\omega_b\delta b^\dag_j\delta b_j+g_j(\delta m_j+\delta m^\dag_j)(\delta b_j+\delta b^\dag_j)
	-i\sum^{N-1}_{j=1}\sum^N_{l=j+1}\sqrt{\kappa_j\kappa_l}e^{i\varphi_{jl}}(\delta m^\dag_j\delta m_l+\delta m_j\delta m^\dag_l),
\end{aligned}
\end{equation}
where $g_j=\tilde{g}_j\langle m_j\rangle$. It is the Hamiltonian in Eq.~\eqref{Hamsystem} in the main text. Here, the coupling strength $g_j$ is assumed to be real-valued for simplicity and without loss of generality. Furthermore, for notational convenience, we adopt the convention $\delta o\to o, o=m_j,b_j$ and $\tilde{\Delta}_j\to\Delta_j$ throughout the main text and subsequent analysis.

\section{Results for Remote Multimode Magnon-Phonon Entanglement}\label{appc}
This appendix provides the system Hamiltonians and drift matrices for the models depicted in Figs.~\ref{Multient}(a) and~\ref{Multient}(b) of Sec.~\ref{Multimode}. We first consider the hybrid model presented in Fig.~\ref{Multient}(a), which is designed to generate remote entanglement between a single phonon and multiple magnons. The system parameters are chosen as $g_1=g$, $g_2=g_3=\cdots g_N=0$, $\varphi_{12}=\varphi$, $\varphi_{23}=\varphi_{34}=\cdots=\varphi_{(N-1)N}=2\pi$. The detunings and radiative decay rates of all magnon modes are assumed to be identical, i.e.,  $\Delta_1=\Delta_2=\cdots=\Delta_N=\Delta$ and $\kappa_1=\kappa_2=\cdots \kappa_N=\kappa$. Under these conditions, the system Hamiltonian in Eq.~\eqref{Hamsystem} turns into
\begin{equation}\label{Hamapho}
	\begin{aligned}
	H&=\sum^N_{j=1}(\Delta m^\dag_jm_j+\omega_bb^\dag_jb_j)+
	g(m_1+m^\dag_1)(b_1+b^\dag_1)-i\sum^N_{j=2}\kappa e^{i\varphi}(m^\dag_1m_j+m_1m^\dag_j)-i\kappa\sum^{N-1}_{j=2}\sum^{N}_{l=j+1}(m^\dag_jm_l+m_jm^\dag_l).
	\end{aligned}
\end{equation} 
Under this condition, the operator vector $u(t)$ in Eq.~\eqref{dotuwhole} can be simplified to
$u(t)=[X_{m_1},Y_{m_1},X_{b_1},Y_{b_1},X_{m_2},Y_{m_2},X_{m_3},Y_{m_3},\cdots X_{m_N},Y_{m_N}]^T$, and the corresponding drift matrix $A$ in Eq.~\eqref{dotuwhole} turns into
\begin{equation}\label{Adriftapho}
\begin{aligned}
A&=\begin{bmatrix}
-\tilde{\kappa} & \Delta & 0 & 0 & -\kappa\cos\varphi & \kappa\sin\varphi &-\kappa\cos\varphi & \kappa\sin\varphi &\cdots &-\kappa\cos\varphi & \kappa\sin\varphi\\
-\Delta & -\tilde{\kappa} & -2g & 0 &-\kappa\sin\varphi &-\kappa\cos\varphi &-\kappa\sin\varphi &-\kappa\cos\varphi &\cdots&-\kappa\sin\varphi &-\kappa\cos\varphi\\
0 & 0 & -\kappa_b & \omega_b & 0 & 0 & 0 & 0 &\cdots & 0 & 0\\
-2g & 0 & -\omega_b & -\kappa_b & 0 & 0 & 0 & 0 &\cdots & 0 & 0\\
-\kappa\cos\varphi & \kappa\sin\varphi & 0 & 0 & -\tilde{\kappa} &\Delta & -\kappa& 0 &\cdots &-\kappa & 0\\
-\kappa\sin\varphi & -\kappa\cos\varphi & 0 & 0 &-\Delta & -\tilde{\kappa} & 0 & -\kappa & \cdots &0 &-\kappa\\
-\kappa\cos\varphi & \kappa\sin\varphi & 0 & 0 & -\kappa & 0 & -\tilde{\kappa} & \Delta & \cdots &-\kappa & 0\\
-\kappa\sin\varphi & -\kappa\cos\varphi & 0 & 0 & 0 & -\kappa &-\Delta & -\tilde{\kappa} & \cdots &0 &-\kappa\\
\vdots & \vdots & \vdots & \vdots & \vdots & \vdots & \vdots & \vdots & \ddots & \vdots &\vdots\\
-\kappa\cos\varphi &\kappa\sin\varphi & 0 & 0 & -\kappa & 0 &-\kappa & 0 &\cdots & -\tilde{\kappa} & \Delta\\
-\kappa\sin\varphi & -\kappa\cos\varphi & 0 & 0 & 0 & -\kappa & 0 &-\kappa &\cdots &-\Delta & -\tilde{\kappa}
\end{bmatrix}
\end{aligned}
\end{equation}
with dimensions $(2N+2)\times(2N+2)$. $\tilde{\kappa}$ is defined as the sum of the radiative decay rate $\kappa$ and the intrinsic decay rate $\gamma$ of magnon mode, i.e., $\tilde{\kappa}\equiv\kappa+\gamma$. Subsequently, by utilizing the system dynamics described in Eq.~\eqref{cmvt}, one can numerically calculate the CM $V(t)$ for this hybrid system. Moreover, by applying the entanglement criterion in Eq.~\eqref{lognegayivity}, the dynamical evolution of entanglement can be obtained, as depicted in Fig.~\ref{Multient}(c) and (e) of the main manuscript.

\begin{figure}[htbp]
	\centering
	\includegraphics[width=0.32\textwidth]{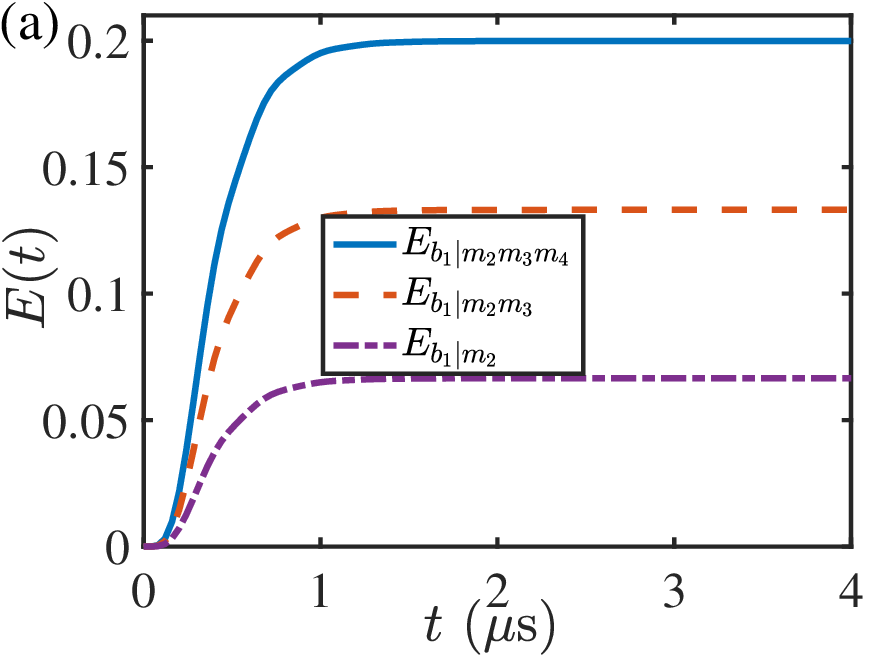}
	\includegraphics[width=0.32\textwidth]{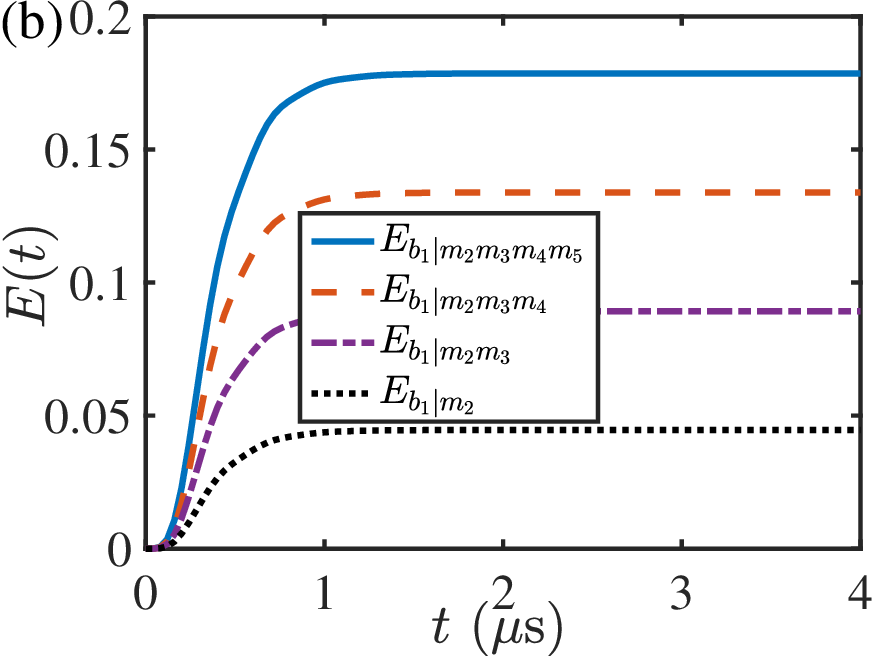}
	\includegraphics[width=0.32\textwidth]{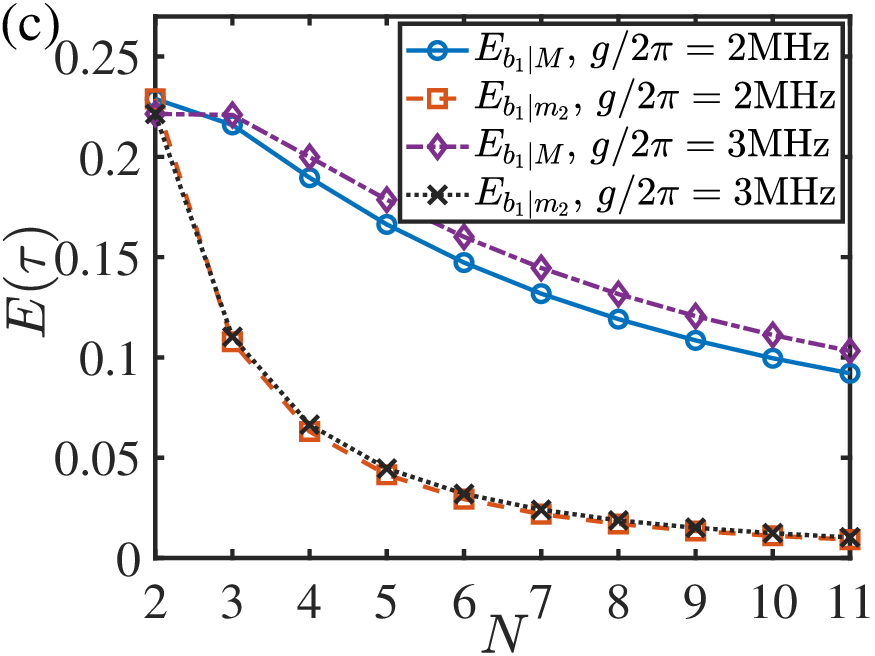}
	\caption{[(a), (b)] Time evolution of the LNs $E(t)$ for various bipartitions with $N=4$ and $N=5$, respectively. The coupling strength is set to $g/2\pi=2$~MHz. (c) The LNs $E_{b_1|M}\equiv E_{b_1|m_2\cdots m_N}$ and $E_{b_1|m_2}$ evaluated at time $\tau=4~\mu$s as a function of the number of YIGs $N$. Here, the other parameters are set to $\varphi=\pi$, $\Delta=-\omega_b$, $\omega_b/2\pi=10$ MHz, $\epsilon/2\pi=10$ GHz, $\kappa_b/2\pi=100$ Hz, $\kappa/2\pi=3$ MHz, $\gamma/2\pi=1$ MHz, and $T=10$ mK.}\label{appcfig}
\end{figure}

In Fig.~\ref{appcfig}(a), we present the time evolution of the LNs $E_{b_1|m_2m_3m_4}$, $E_{b_1|m_2m_3}$, and $E_{b_1|m_2}$ for a system with $N=4$ YIGs. All quantities converge to steady values, satisfying the relations $E_{b_1|m_2m_3}=2E_{b_1|m_2}$ and $E_{b_1|m_2m_3m_4}=3E_{b_1|m_2}$. Similarly, Fig.~\ref{appcfig}(b) shows the corresponding results for $N=5$, where the steady values obey $E_{b_1|m_2m_3m_4m_5}=4E_{b_1|m_2}$ and $E_{b_1|m_2m_3m_4}=3E_{b_1|m_2}$. Figure~\ref{appcfig}(c) shows the dependence of $E_{b_1|M}$ and $E_{b_1|m_2}$ at time $\tau$ on the number of magnon modes $N$. Independent of the coupling strength $g$, the results clearly satisfy the relation $E_{b_1|M}=(N-1)E_{b_1|m_2}$, where $E_{b_1|M}\equiv E_{b_1|m_2\cdots m_N}$. Moreover, the imposed phase relations ensure that $E_{b_1|m_2}=E_{b_1|m_3}=\cdots=E_{b_1|m_N}$. We thus conclude that the entanglement between a single phonon and multiple distant magnons equals the sum of the individual phonon-magnon entanglements. Furthermore, additional numerical analysis confirms that $E_{m_j|m_l}=0$ for all $j,l=2,\dots,N$ with $j\neq l$, indicating the absence of entanglement between magnon modes.

\begin{figure}[htbp]
	\centering
	\includegraphics[width=0.32\textwidth]{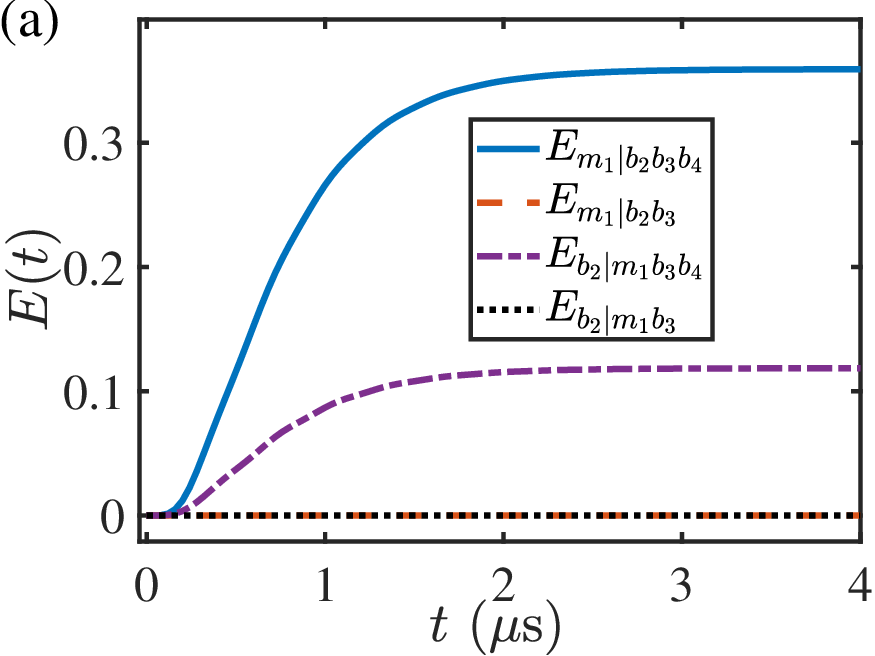}
	\includegraphics[width=0.32\textwidth]{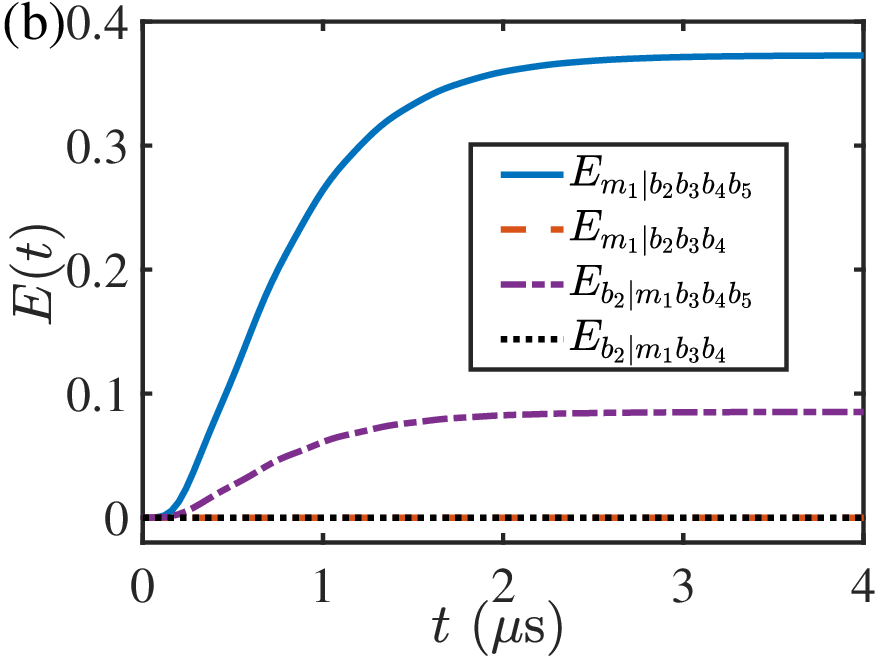}
	\includegraphics[width=0.32\textwidth]{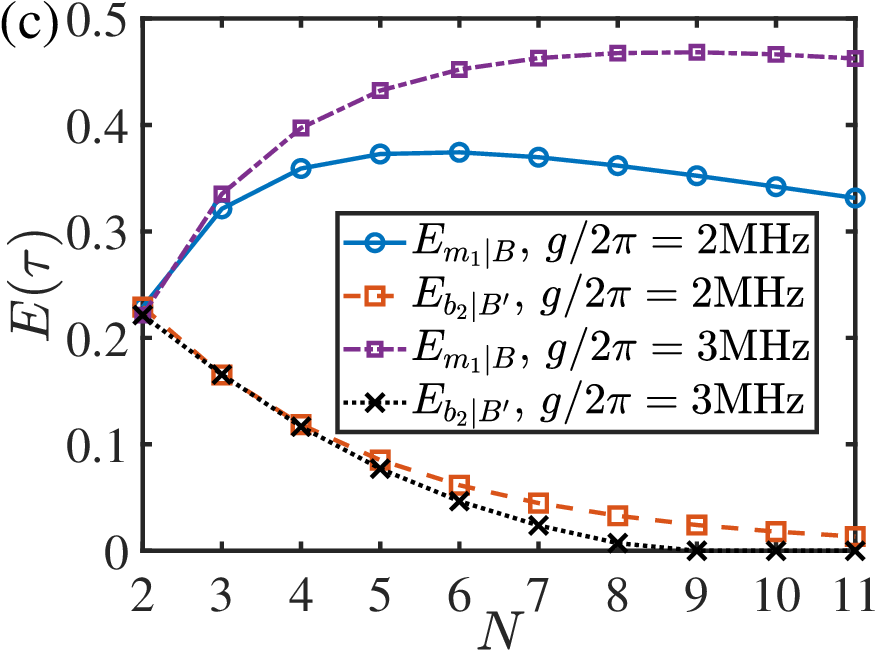}
	\caption{[(a), (b)] Time evolution of the LNs $E(t)$ for various bipartitions with $N=4$ and $N=5$, respectively. The coupling strength is set to $g/2\pi=2$~MHz. (c) LNs $E_{m_1|B}\equiv E_{m_1|b_2b_3\cdots b_N}$ and $E_{b_2|B'}\equiv E_{b_2|m_1b_3\cdots b_N}$ evaluated at time $\tau=4~\mu$s as a function of the number of YIGs $N$.  Here, all other parameters are identical to those in Fig.~\ref{appcfig}.}\label{appcfig2}
\end{figure}

Next, we consider the hybrid model presented in Fig.~\ref{Multient}(b), which is designed to generate remote entanglement between a single magnon and multiple phonons. The system parameters are chosen as $g_1=0$, $g_2=g_3=\cdots =g_N=g$, $\varphi_{12}=\varphi$, $\varphi_{23}=\varphi_{34}=\cdots=\varphi_{(N-1)N}=2\pi$. Additionally, the detunings and decay rates of all magnon modes are assumed to be identical, namely, $\Delta_1=\Delta_2=\cdots=\Delta_N=\Delta$ and $\kappa_1=\kappa_2=\cdots \kappa_N=\kappa$. Under these conditions, the system Hamiltonian in Eq.~\eqref{Hamsystem} simplifies into
\begin{equation}\label{Hamamag}
\begin{aligned}
	H&=\sum^N_{j=1}(\Delta m^\dag_jm_j+\omega_bb^\dag_jb_j)+
	\sum^N_{j=2}[g(m_j+m^\dag_j)(b_j+b^\dag_j)-i\kappa e^{i\varphi}(m^\dag_1m_j+m_1m^\dag_j)]-i\kappa\sum^{N-1}_{j=2}\sum^{N}_{l=j+1}(m^\dag_jm_l+m_jm^\dag_l).
\end{aligned}
\end{equation}

Under this case, the operator vector $u(t)$ in Eq.~\eqref{dotuwhole} reduces into $u(t)=[X_{m_1},Y_{m_1},X_{m_2},Y_{m_2},X_{b_2},Y_{b_2},\cdots X_{m_N},Y_{m_N},X_{b_N},Y_{b_N},]^T$, and the corresponding drift matrix $A$ in Eq.~\eqref{dotuwhole} takes the form
\begin{equation}\label{Adriftamag}
\begin{aligned}
A&=\begin{bmatrix}
	-\tilde{\kappa} & \Delta & -\kappa\cos\varphi & \kappa\sin\varphi & 0 & 0 & \cdots &-\kappa\cos\varphi & \kappa\sin\varphi & 0 & 0\\
	-\Delta & -\tilde{\kappa} & -\kappa\sin\varphi &-\kappa\cos\varphi &0 & 0 & \cdots&-\kappa\sin\varphi &-\kappa\cos\varphi &0 &0\\
	-\kappa\cos\varphi & \kappa\sin\varphi & -\tilde{\kappa} &\Delta & 0 & 0 & \cdots&-\kappa& 0  &0 & 0\\
	-\kappa\sin\varphi & -\kappa\cos\varphi &-\Delta & -\tilde{\kappa} & -2g & 0 &  \cdots&0 &-\kappa & 0 &0\\
	0& 0 & 0 & 0 & -\kappa_b & \omega_b & \cdots &0 & 0 &0 & 0\\
	0 & 0 &-2g & 0 & -\omega_b & -\kappa_b & \cdots &0 &0 &0 & 0\\
	\vdots & \vdots & \vdots & \vdots & \vdots & \vdots & \vdots & \vdots & \ddots & \vdots &\vdots\\
	-\kappa\cos\varphi &\kappa\sin\varphi & -\kappa & 0 & 0 & 0 &\cdots & -\tilde{\kappa} & \Delta & 0 & 0\\
	-\kappa\sin\varphi & -\kappa\cos\varphi & 0 & -\kappa& 0 & 0 &\cdots &-\Delta & -\tilde{\kappa} &  -2g & 0\\ 
	0 & 0 & 0 & 0 & 0 &0 & \cdots & 0 & 0 & -\kappa_b &\omega_b\\
	0 & 0 & 0 & 0 & 0 &0 & \cdots & -2g & 0 &-\omega_b &-\kappa_b
\end{bmatrix}
\end{aligned}
\end{equation}
with dimensions $(4N-2)\times(4N-2)$. Subsequently, by utilizing the system dynamics described in Eq.~\eqref{cmvt}, one can numerically calculate the CM $V(t)$ for the hybrid system. Moreover, by applying the entanglement criterion in Eq.~\eqref{lognegayivity}, the dynamical evolution of entanglement can be derived, as depicted in Fig.~\ref{Multient}(d) and (f) of the main text.

Figure~\ref{appcfig2}(a) and (b) present the time evolution of various LNs for systems comprising $N=4$ and $N=5$ YIGs, respectively. For a fixed $N$, the LN between any single mode and the remainder of the system converges to a steady value after sufficiently long evolution. The other LNs, not shown in the figure, can be readily inferred from the previously established phase relations. Notably, the entanglement between a single magnon and the collective phonon modes (blue solid line) is considerably stronger than that between a single phonon and all other modes combined (red dashed line). In contrast, when the multipartite partition excludes any of the remaining modes, the corresponding entanglement remains identically zero throughout the entire evolution (purple dash-dotted and black dotted lines). This feature closely resembles that of a GHZ state, in which tracing out any individual subsystem completely destroys all entanglement, emphasizing the delicate and global nature of the quantum correlations in the system. In Fig.~\ref{appcfig2}(c), the LNs at time $\tau$ are plotted as functions of the total number of YIG spheres $N$. Here, $E_{m_1|B}\equiv E_{m_1|b_2b_3\cdots b_N}$ and $E_{b_2|B'}\equiv E_{b_2|m_2b_3\cdots b_N}$. It is observed that a stronger magnomechanical coupling $g$ enhances $E_{m_1|B}$ while reducing $E_{b_2|B'}$ for a fixed $N$. Moreover, $E_{b_2|B'}$ decreases monotonically with increasing $N$. In particular, for a larger coupling strength $g/2\pi=3$MHz, $E_{b_2|B'}$ vanishes when $N\ge9$, indicating the absence of genuine multimode entanglement.
\end{widetext}

\bibliographystyle{apsrevlong}
\bibliography{reference}

\end{document}